\documentclass[fleqn,usenatbib]{mnras}

\usepackage{newtxtext,newtxmath}
\usepackage[T1]{fontenc}

\DeclareRobustCommand{\VAN}[3]{#2}
\let\VANthebibliography\thebibliography
\def\thebibliography{\DeclareRobustCommand{\VAN}[3]{##3}\VANthebibliography}

\usepackage{graphicx}	% Including figure files
\usepackage{amsmath}	% Advanced maths commands
\usepackage{tabularx}
\usepackage{float}

\title[Source effects on subhalo detectability]{Impacts of source morphology on the detectability of subhalos in strong lenses.}

\author[T. J. Hughes et al.]{
Tyler J. Hughes$^{1,2}$,
Karl Glazebrook$^{1}$,
Colin Jacobs$^{1}$
\\
$^{1}$Swinburne University of Technology, John st, Melbourne 3122, Australia\\
$^{2}$ARC Centre of Excellence for Dark Matter Particle Physics
}

\date{Accepted XXX. Received YYY; in original form ZZZ}

\pubyear{2023}

\begin{document}
\label{firstpage}
\pagerange{\pageref{firstpage}--\pageref{lastpage}}
\maketitle

\begin{abstract}
We provide an analysis of a convolutional neural network's ability to identify the lensing signal of single dark matter subhalos in strong galaxy-galaxy lenses in the presence of increasingly complex source light morphology. We simulate a balanced dataset of 800,000 strong lens images both perturbed and unperturbed by a single subhalo ranging in virial mass between $10^{7.5} M_{\odot} - 10^{11}M_{\odot}$ and characterise the source complexity by the number of Sersic clumps present in the source plane ranging from 1 to 5. Using the ResNet50 architecture we train the network to classify images as either perturbed or unperturbed. We find that the network is able to detect subhalos at low enough masses to distinguish between dark matter models even with complex sources and that source complexity has little impact on the accuracy beyond 3 clumps. The model was more confident in its classification when the clumps in the source were compact, but cared little about their spatial distribution. We also tested for the resolution of the data, finding that even in conditions akin to natural seeing the model was still able to achieve an accuracy of 74\% in our highest peak signal-to-noise datasets, though this is heavily dominated by the high mass subhalos. It's robustness against resolution is attributed to the model learning the flux ratio anomalies in the perturbed lenses which are conserved in the lower resolution data.

%We find that beyond 3 clumps the source complexity has a minimal impact on model accuracy likely attributed to ResNets use of the multiple images in the lens to identify uncorrelated features in the source. The model has an easier time finding the subhalo's lensing signal when the clumps in the source are concentrated and their distribution is compact, which suggests that an ideal source will be large, as to probe a significant portion of the lens, but with a sizeable amount of fine structure. 

\end{abstract}


%%%%%%%%%%%%%%%%%%%%%%%%%%%%%%%%%%%%%%%%%%%%%%%%%%

%%%%%%%%%%%%%%%%% BODY OF PAPER %%%%%%%%%%%%%%%%%%

\hypertarget{introduction}{%
\section{Introduction}\label{introduction}}

Dark matter dominates the matter density of the universe and as such drives structure formation \citep{2012AnP...524..507F}. The physical properties of dark matter influence structure formation, and so cosmological observations constrain potential dark matter models or candidates. One theory \citep{1991NuPhB.360..145G} postulates dark matter
as a thermal relic born from quantum fluctuations in the early universe
that seeded regions of overdensity in the matter distribution, which have
evolved as the universe has expanded into the galaxy scale and galaxy
cluster scale dark matter halos that we see today. Overdensities smaller
than the free streaming length of the dark matter particle are unable to
survive, introducing a characteristic scale into the matter power
spectrum below which structure formation is suppressed \citep{1974A&A....37..225M,2014ApJ...788...27I,2013MNRAS.433.1573S}. The time at
which the early velocities of dark matter particles become
non-relativistic depends on their mass: more massive particles transition much sooner than lighter particles resulting in much
shorter free streaming lengths. This led to a classification model of
different dark matter candidates which identified the particle as either
Hot Dark Matter (HDM) denoting lighter particles, Cold Dark Matter (CDM)
pertaining to heavier candidates, or Warm Dark Matter (WDM) for the
intermediate masses. HDM has long been ruled out given its predictions
of large scale structure in the early universe are inconsistent with
observations \citep{1983ApJ...274L...1W}. CDM has stood the tests of time far more rigorously,
better predicting large scale structure than HDM \citep{1985ApJ...292..371D,2006Natur.440.1137S,2005Natur.435..629S}, but also successfully reproduces other physical properties such as the rotation
curves of galaxies. 

Despite CDM's success, there are still other plausible dark matter models. In the context of thermal relic dark matter, a galactic dark matter halo is expected to have accreted a large population of smaller subhalos over time. Consequently, free streaming effects introduce a minimum mass threshold for these subhalos. As a
result the mass distribution of subhalos is a function of the mass of
the dark matter particle. WDM candidates like sterile neutrinos will
then produce very different subhalo populations than CDM which predict
subhalos all the way down to earth masses \citep{10.1093/mnras/stt829}. By measuring the subhalo mass
function (SHMF) which parameterises the subhalo mass distribution you
can distinguish between CDM and other dark matter models such as WDM or
even more exotic models like fuzzy dark matter \citep{PhysRevLett.85.1158}. Observational measurements of sub-galactic dark
matter structure are notoriously difficult to make \citep{1999ApJ...522...82K,2011MNRAS.415L..40B,2011ApJ...742...20W} leaving room for several WDM candidates.
Large scale structure surveys use central galaxies to trace the
positions and masses of large dark matter halos, but there are several
factors that can prevent star formation in subhalos \citep{2016MNRAS.456...85S,2006Natur.442..539M,2002ApJ...580..627W}. This has prompted researchers to look for alternative methods such
as strong gravitational lensing for probing substructure.

The strong lensing of distant galaxies provides a prime opportunity for
substructure detection via their gravitational deflections of light. Where the majority of the lensing effect is a
result of the main dark matter halo and the central galaxy in the lens,
subhalos introduce subtle perturbations into the final lens image.
These perturbations may be identified by measuring the
deviations from a reconstructed image assuming a smooth mass
distribution in the lens model. However, this process is complicated by
the fact that the final image is not only a function of the lensing
potential but also the surface brightness in the source plane \citep{10.1111/j.1365-2966.2005.09523.x}. Previous
studies have attempted to detangle these two attributes in a couple of
different ways. When comparing the multiple images of a lens any
uncorrelated features should be a result of lensing perturbations \citep{1998MNRAS.295..587M, Dalal_2002,Kochanek_2004}, however to some extent these can be accounted for once you consider the main halo's dwarf galaxy population \citep{More_McKean_More_Porcas_Koopmans_Garrett_2009}. Hence a more elaborate process was employed by
\cite{vegetti_bayesian_2009} that iteratively reconstructs both the surface brightness in the source plane and a grid based substructure distribution simultaneously using Bayesian statistics. Given the final substructure distribution the best fitting density profile can
infer the position and mass of the subhalos. Applying this method to HST
data, two dark subhalo detections have been claimed so far, with masses
of {\(3.5 \times 10^{9}M_{\odot}\)} and
{\(1.9 \times 10^{8}M_{\odot}\)} \citep{vegetti_detection_2010,vegetti_gravitational_2012}, though measuring the SHMF is yet to be
achieved. These detections aren't
without challenge, though. Vegetti's method assumes that the entirety of the lensing mass is
within a single lens plane and doesn't take into account that the perturbation may come from a dark matter halo in the line of sight. This type of halo significantly contributes to the overall
effective lensing potential and can be difficult to distinguish from
subhalos in the main lens plane \citep{10.1093/mnras/sty159,PhysRevD.102.063502,2017MNRAS.468.1426L}. This can lead to overestimation of the
subhalo mass fraction in the lens and inaccurate modelling of the SHMF.
\cite{sengul_substructure_2021} claimed that the first of Vegetti's detections is in fact a line of sight halo. By setting the redshift of the perturber as a free parameter they found a better fitting lens model with the main lens at {\(z = 0.881\)} and the perturber at {\(z_{\text{los}} = 1.22_{- 0.11}^{+ 0.11}\)}.

The advent of machine learning has not only given astronomers the means to sort through large amounts of data far more efficiently, but it allows us to measure the properties of systems without
needing to explicitly specify or model the underlying physics. For substructure detection this means that we can in theory measure the lensing signal from the subhalo population without needing to model either the lens or source, avoiding the need to manually disentangle the substructure-source contributions to fine structure in the lensed image. Recent works have
shown that machine learning can be applied in different ways to identify
substructure signals in strong lens systems.
\cite{diaz_rivero_direct_2020} demonstrated that a convolutional neural network could successfully be
used to classify whether or not simulated strong lens images contained
substructure. Performing the analysis both with single subhalos and full
subhalo populations the model was able to accurately identify single
subhalos down to {\(10^{9}M_{\odot}\)} about {\(60\%\)} of the time in images with little noise, but dropped off completely when the noise was above 10\% of the mean signal in the lensed arcs. Using full subhalo populations didn\textquotesingle t do much
to improve the model's sensitivity to subhalo mass, but made it less
susceptible to noise. 

\cite{ostdiek_extracting_2022} have suggested using semantic image
segmentation methods to physically map out the subhalo population in a
lens. Such neural networks classify each pixel as belonging to a particular
class, in this case either the main lens, a subhalo of a given mass, or
the background. This technique performs well in identifying the mass and
position of subhalos larger than {\(10^{9.5}M \odot\)}, but was only able
to detect 17\% of {\(10^{9}M \odot\)} and struggled to detect any
subhalos lower than {\(10^{8}M \odot\)}. These two procedures are
considered as direct detection methods as they try to explicitly infer
the physical properties of the subhalo population.

Recently more
statistical methods have been tested \citep{wagner-carena_hierarchical_2021,wagner-carena_images_2022} which attempt to infer
parameters of the SHMF by using models such as neural posterior
estimators in combination with a hierarchical inference framework. The
neural posterior estimator learns to predict the lens parameters and the normalisation and slope of the subhalo mass function. The
hierarchical inference framework then combines the estimates to infer a
universal subhalo mass function. Once the network was trained this method was able to recover the SHMF
normalisation consistent with dark matter-only simulations to within an
order of magnitude using only 100 lenses. The
study developed datasets that attempted to replicate realistic HST data by
including noise, data processing effects, and complex source
morphologies from the COSMOS survey \citep{2007ApJS..172..196K}.

Given a WDM SHMF will generally begin to significantly deviate from CDM below
{\(10^{8}M_{\odot}\)}, these studies provide good evidence that given further refinement machine
learning could be sensitive enough to identify subhalos in the mass range
needed to distinguish between the two, but there are still several
questions that need to be answered before these models can be applied to
real data. One such question, and the focus of this work, is how exactly
source morphology influences a model's ability to identify substructure.
Though machine learning requires no prior knowledge of the source to
identify a subhalo signal, that doesn\textquotesingle t mean that
it\textquotesingle s immune to source-substructure degeneracy and it may
be that when applied to real images of galaxy-galaxy lenses the model
assumes more low mass substructure than is actually present. The
works mentioned previously do attempt to build complexity into the
source morphology to differing degrees, either in the form of multiple
clumps, or using real images of COSMOS galaxies, but none explicitly
explore how the morphology impacts their model's performance.

In this paper we explore a convolutional neural network's ability to
identify the lensing signal of single subhalos in simulated images of
galaxy-galaxy strong lenses and by gradually building complexity into
the source, measure how source morphology impacts its ability to
identify substructure. With JWST data now becoming increasingly available we also wish to explore the relationship between model performance and the resolution of the data.
In section 2 we outline the methods we use to
generate the dataset of simulated images. In section 3 we then present
the architecture for our convolutional neural network and the process we
used to train and evaluate it. Then in section 4 we evaluate the results
and discuss our finding, and finally summarise our main conclusions in
section 5.

\hypertarget{simulation-methods}{%
\section{Simulation Methods}\label{simulation-methods}}

\begin{figure*}
    \centering
    \includegraphics[width=0.8\linewidth]{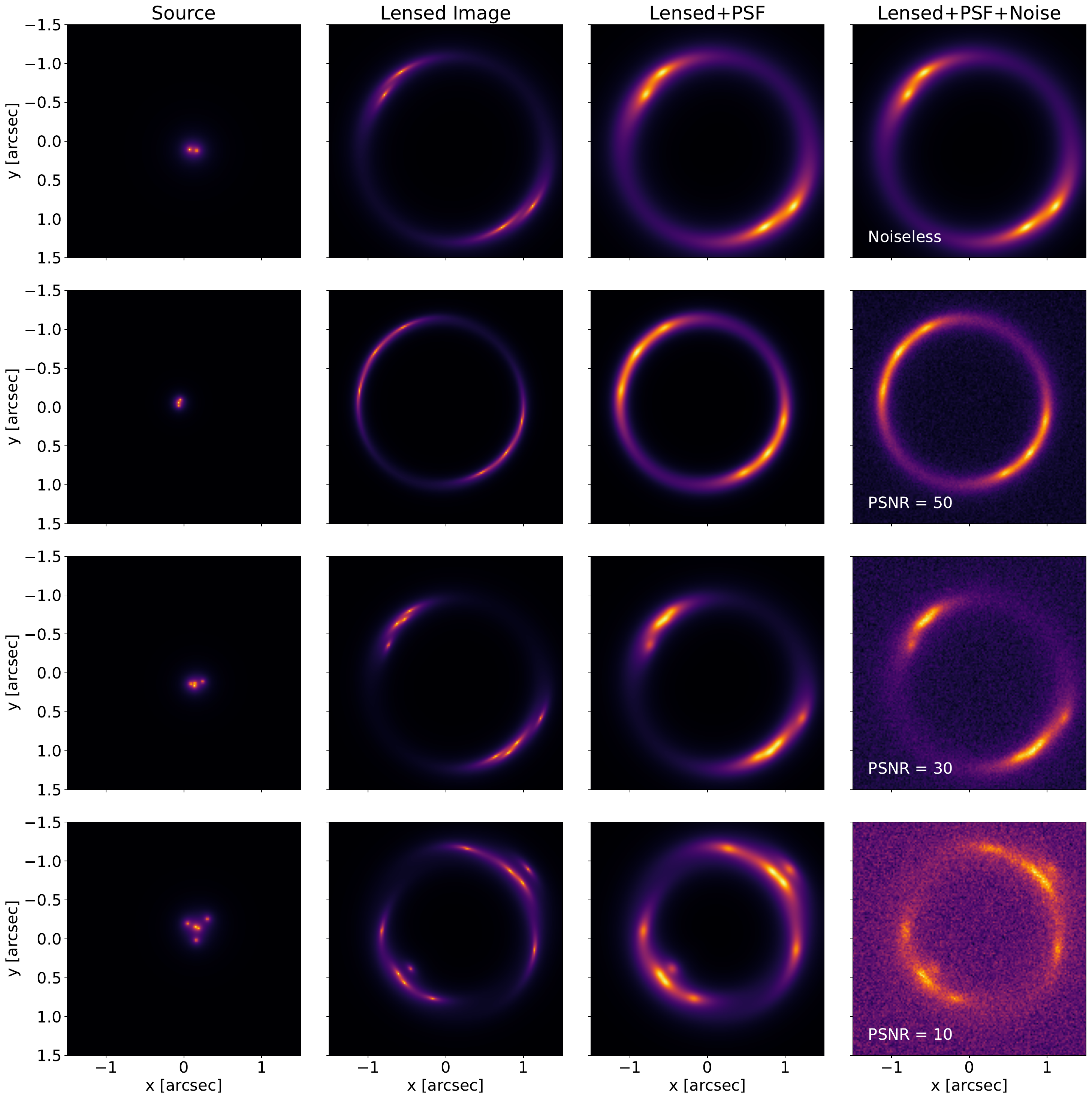}
    \caption{Example of the simulation pipeline for 4 different source configurations. From top to bottom, images in each row have an increasing number of clumps in the source ranging from two in the top example to 5 at the bottom.}
    \label{fig:sim_pipeline}
\end{figure*}

The purpose of this work is to develop a CNN that can identify whether
or not a subhalo is present in an image of a strong galaxy-galaxy lens
and explore how the morphology of the source impacts the CNN\textquotesingle s
ability to identify the subhalo's signal. This requires a supervised
learning approach in which a dataset's labels are already known. The only
way to ensure this is to use a simulated dataset in which there is
explicit control over the presence and properties of the subhalo. We use the
python based software \textsc{DeepLenstronomy}
\citep{deeplenstronomy} which builds on top of the
lens modelling library \textsc{Lenstronomy} \citep{Birrer2021} to generate the simulated
images but allows us to define parameter distributions and dataset
specifications using simple \textsc{YAML}\footnote{https://yaml.org} files, making it easier to build full
datasets for machine learning purposes. For the subhalos we use
another python library called \textsc{PyHalo} \citep{PyHalo} which has a comprehensive range of functionality to define subhalo populations in a format that can be directly input into a \textsc{Lenstronomy} lens model. It allows us to build our subhalos using only the virial mass before infall into its parent halo. The simulation pipeline is as follows. A light profile is defined to model the surface brightness in the source plane, the source is lensed by a deflector with some mass defined by some density profile, the resulting lensed image is convolved with a Gaussian point spread function, then noise is added to produce the final image. An example of the pipeline can be seen in figure \ref{fig:sim_pipeline} and the details of each component are described in the remainder of this section.\\

\hypertarget{source-model}{%
\subsection{Source Model}\label{source-model}}

For the purposes of this study we focus on using a simple parameterised model to define
the surface brightness in the source plane which will allow us to
explicitly explore the relationship between the parameters of the source and model performance. Sources are uniformly positioned between redshift 1 and 2 which is consistent with the source redshifts found in the ASTRO 3D Galaxy Evolution with Lenses (AGEL) survey \citep{Tran_2022}. In this redshift range we expect source galaxies to be diverse in their morphologies, forming spheroidal, disky, and irregular structures such as in mergers \citep{GUANWEN201820}. To properly understand the impact of these morphologies on CNN performance it's important to clearly define
the parameters of our source model. We divide the definition of morphology into two
components; complexity and geometry. Complexity refers to the amount by
which the source deviates from a smooth distribution (the clumpiness
of the source), while geometry refers to the shape of both the individual
clumps and the whole source overall. We build the source as follows. We
define the simplest source model in our dataset as a single Sersic
profile \citep{1968adga.book.....S},
{\[I(R) = I_{0}\exp\left\lbrack - b_{n}(R/R_{s})^{\frac{1}{n}} \right\rbrack\]}where
{\(I_{0}\)} is the magnitude, {\(R_{1/2}\)} is the radius within which
half of the total light is found, {\(n\)} is the Sersic index which
defines the steepness of the distribution and {\(b_{n}\)} is a parameter
dependent on {\(n\)} given by the approximation: {\[b_{n} \approx 1.999n - 0.327\]}We can
then define the shape of the single clump using {\(R_{1/2}\)} and {\(n\)},
both of which are a measure of the concentration of the light. To build
complexity into the model we introduce additional Sersic profiles into
the source plane making parameterisation simply the total number of
clumps. For any given source with multiple clumps we give each clump the
same Sersic parameters so that we can draw more direct conclusions about
model performance against {\(R_{1/2}\)} and {\(n\)}. When distributing the clumps in the source plane we define the position of a
central clump and the remaining clump positions are drawn from an
elliptical-exponential radial distribution. This gives us two extra
geometric parameters; {\(\lambda\)} defining the concentration of the
distribution and {\(e\)} denoting the ellipticity. The magnitude of the
central clump is arbitrarily set to 20. The exact number is not consequential given there is no deflector light and the images are mean normalised before the network sees them. It's only needed to parameterise the flux of the remaining clumps which are taken from a normal distribution with a mean
equal to the magnitude of the central clump and standard deviation of 0.1. Distributions for each of these parameters can be seen in Table \ref{tab:source_distribution}.

\begin{table}
    \centering
    \begin{tabularx}{0.4\textwidth}{X l}
    \hline
        \textbf{Component} & \textbf{Distribution}\\
        \hline
        \textbf{\textit{Source central clump}}\\
        Half-light radius (")  & $R_{\rm S} \sim \mathcal{U}(0.1,0.5)$  \\
        Sersic index  & $n \sim \mathcal{U}(1,4)$\\
        Apparent magnitude & $m = 20$\\
        x position (") & $x \sim \mathcal{U}(-0.2,0.2)$\\
        y position (")& $x \sim \mathcal{U}(-0.2,0.2)$\\
        redshift & $z_{\text{source}} \sim \mathcal{U}(1,2)$\\
        \hline
        \textbf{\textit{Source scattered clumps}}\\
        Half-light radius (")  & $R_{\rm S,clump}=R_{S}$  \\
        Sersic index  & $n_{\rm clump}=n$\\
        Scale parameter (") & $\lambda = 2*\mathcal{U}(0.1,0.5)$\\
        Apparent magnitude & $m_{\rm clumps} \sim \mathcal{N}(m, 0.1)$\\
        x position (") & $x \sim x_{\text{source}}+\mathcal{E}(\lambda)$\\
        y position (")& $y \sim y_{\text{source}}+\mathcal{E}(\lambda)$\\
        redshift & $z_{\text{source}} \sim \mathcal{U}(1,2)$\\
        \hline
        \textbf{\textit{Main Lens}}\\
        Mass & $M_{200}=10^{13}M_{\odot}$\\
        redshift & $z_{\text{lens}} \sim \mathcal{U}(0.1,0.5)$\\
        x position (") & $x=0$\\
        y position (")& $y=0$\\
        
    \end{tabularx} 
    \caption{Distributions for parameters in the source and lens plane. The selection of subhalo parameters is discussed in section 2.3.}
    \label{tab:source_distribution}
\end{table}

\hypertarget{main-lens}{%
\subsection{Main Lens}\label{main-lens}}

The mass distribution of a lensing galaxy can be expressed as a linear
combination of four seperate components: the main dark matter halo, the
baryonic matter of the central galaxy, the dark matter substructure, and
any baryonic matter present in the subhalos. The central galaxy and main
halo can either be treated individually, typically by combining a
Navarro-Frenk-White (NFW; \citealp{navarro_structure_1996}) density profile for the halo and an
Hernquist profile for the baryonic matter, or singularly with a singular
isothermal sphere (SIS) which is indiscriminate of the type of matter
and provides a reasonable approximation to the combined density profile.
The latter approach can be more convenient for lens modelling given
it can be defined by the Einstein radius. For this reason the benefit of the SIS profile is that
the Einstein radius can easily be expressed in terms of {\(M_{200}\)}, the mass contained within a radius {\(R_{200}\)} in which the mean density is 200 times the critical density of the universe, and the distances to the lens and source,
{\[\theta_{E} = \frac{4GM_{200}}{c^{2}R_{200}}\frac{D_{ds}}{D_{s}}\]}
where
{\(D_{s}\)} is the angular diameter distance to the source, and {\(D_{ds}\)}
is the angular diameter distance from the lens to the source. Redshifts for the lens and source
planes are sampled from a uniform distribution ranging from
{\(0.1 < z_{\text{lens}} < 0.5\)} for the lens and {\(1 < z_{\text{source}} < 2\)} for
the source. The velocity dispersion of early type galaxy deflectors tend to fall within {\(158 \text{km/s} < \sigma_v < 220 \text{km/s}\)} \citep{Davis_Huterer_Krauss_2003} which, assuming an SIS, translates to a virial mass to the order of {\(10^{13}M_{\odot}\)}, and so we fix the mass of the deflector at this value.

\hypertarget{subhalo}{%
\subsection{Subhalo}\label{subhalo}}

In this work we wish to explore the more fundamental effects of source
morphology against a single subhalo and leave its
influence on the lensing signal of the full subhalo population for future work. We use a truncated NFW (TNFW) density profile \citep{2009JCAP...01..015B} which is consistent with their mass distribution in cosmological simulations:

\[\rho(r) = \frac{r_{\rm trunc}^{2}}{r^{2} + r_{\rm trunc}^{2}}\frac{\rho_{0}}{r/R_{s}(1 + r/R_{s})^{2}}\]

where {\(r_{\rm trunc}\)} is the truncation radius and accounts for mass lost to tidal disruption, {\(R_{s}\)} is the scale
radius, and {\(\rho_{0}\)} is a characteristic density. The scale radius
can be expressed in terms of the virial radius {\(R_{200}\)} and a
concentration parameter {\(c\)}, while the {\(\rho_{0}\)} can be
calculated using the concentration and redshift of the lens. As a
result, the subhalo density profile can be completely described by the
physical properties {\(M_{\rm sub}\)}, {\(c\)}, {\(z_{\rm lens}\)}, and
{\(r_{\rm trunc}\)}. We assume the subhalo does not host any baryonic matter
and therefore is centrally cuspy. The position of the subhalo in the
lensing plane is chosen such that the flux at that point
is at least {\(50\%\)} of the maximum value. This ensures that it has
enough freedom not to be directly placed on top of an image in the
lensed plane where the signal might be strongest, but still has enough flux to produce a sizeable perturbation. The concentration is drawn from the Diemer
\citep{2019ApJ...871..168D} concentration-mass relation with added scatter.
The truncation radius depends on where the
subhalo is positioned within the parent halo. The closer the subhalo is
to the centre of the main halo the greater the tidal forces and the more
mass is lost to tidal stripping:

\[r_{\text{trunc}} = 1.4\left( \frac{M_{200}}{10^{7}M_{\odot}} \right)^{\frac{1}{3}}\left( \frac{r_{\rm 3D}}{50\text{kpc}} \right)^{\frac{2}{3}} \text{kpc}\]

Where {\(r_{\text{3D}}\)} is the three dimensional distance of the
subhalo to the centre of the main lens. This relation assumes tidal
stripping consistent with the Roche criterion for diffuse masses in an
approximately isothermal mass. {\(r_{\text{3D}}\)} is positioned randomly
within the virial radius of the parent halo. By adding a subhalo to the
main halo we increase the total mass of the system. To correct for this
we include a negative mass sheet in the lens model. Failing to correct
for the mass offset could mean that the model could learn to identify
lenses with a subhalo purely from increase in Einstein radius. This
effect would be particularly evident for large subhalo masses. We
generate dataset mixed subhalo masses
ranging from {\(10^{7.5}M_{\odot}\)} to {\(10^{11}M_{\odot}\)}.

\hypertarget{instrument}{%
\subsection{Data specifications}}

The images are 150x150 pixels and are convolved with an HST-consistent
point spread function (PSF) with a full width half maximum (FWHM) of 0.08". To ensure that detection of the
substructure signal is a resolution limited problem we use a pixel scale
of 0.02" such that the PSF is well sampled. It\textquotesingle s
important to note that the pixel scale in our images are half that of
the usual HST image, however this is often improved through drizzling \citep{2002PASP..114..144F}
and therefore our set value is not so unrealistic. However we do not use correlated noise found between pixels in a drizzled image, which can be mistaken as substructure by the network \citep{diaz_rivero_direct_2020}. We use Gaussian noise and define the amount by the peak signal-to-noise ratio (PSNR): 

$$\text{PSNR} = \frac{I_{\text{peak}}}{\sigma}$$

where $I_{\text{peak}}$ is the maximum signal in the image, and $\sigma$ is the standard deviation of the noise. 

For the resolution experiment we use the exact same dataset as in the HST-like images and simply change the FWHM of the PSF. We choose two wider PSF's with FWHM of 0.5" and 0.17" to reflect natural seeing, and Euclid respectively, and one narrower FWHM of 0.03" akin to the best resolution JWST data. With only the resolution changing between datasets the total flux in the images is consistent, and so to preserve the total signal to noise ratio in a given image we use the same noise used in the HST-like dataset. For consistency we still quantify the noise in these datasets using PSNR, but throughout this paper we always discuss PSNR with respect to the HST-like images. Example images are shown in figure \ref{fig:resolution_sample}.

\hypertarget{cnn-development}{%
\section{Model architecture and training}\label{cnn-development}}

\begin{figure}
\centering
\includegraphics[width=\linewidth]{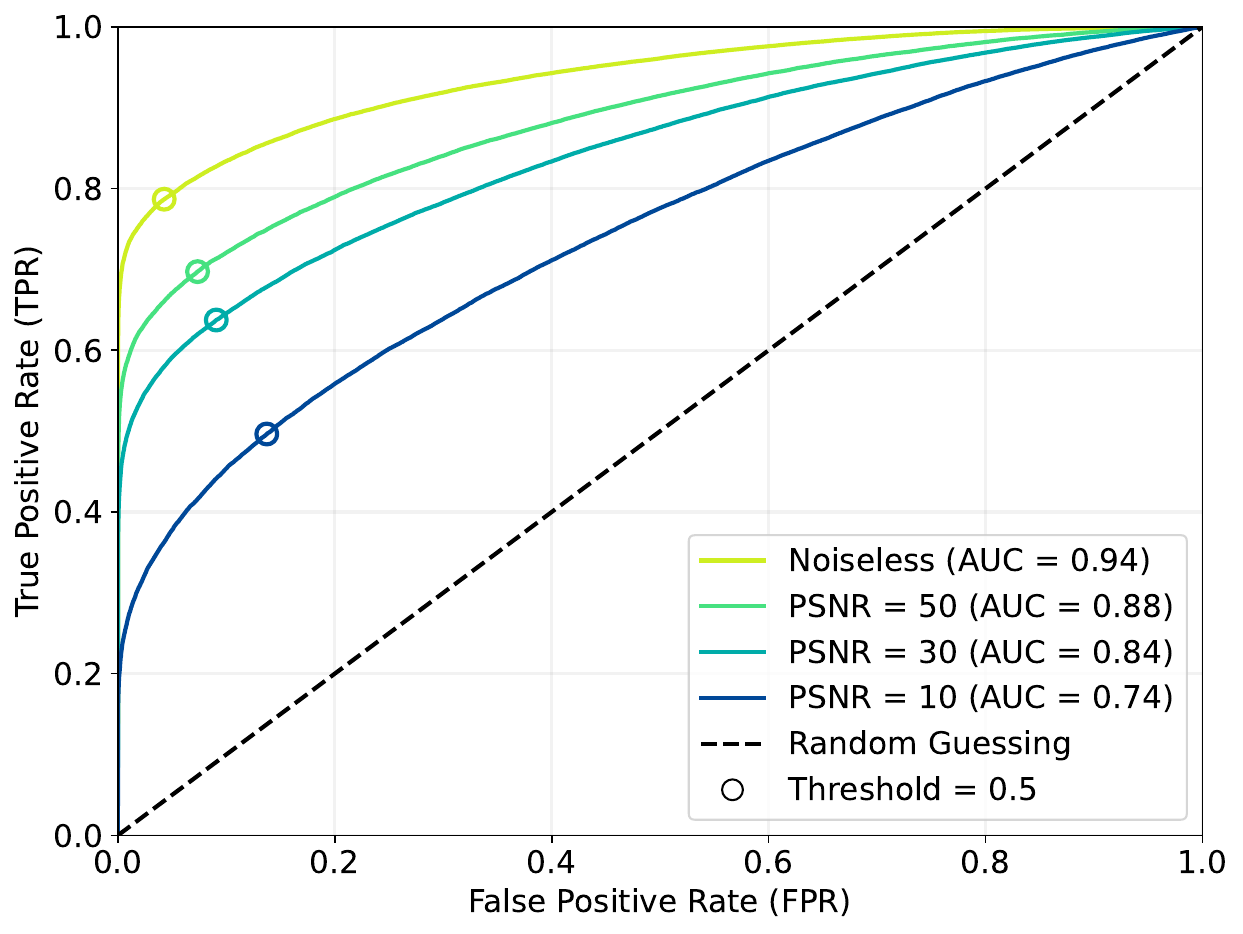}
\caption{ROC's for ResNet50 on the test set of simulated lenses for all four peak signal-to-noise ratios.}
\label{fig:ROC}
\hfill  % fill the gap between the two minipages % adjust the width of the minipage asded
\centering
\includegraphics[width=\linewidth]{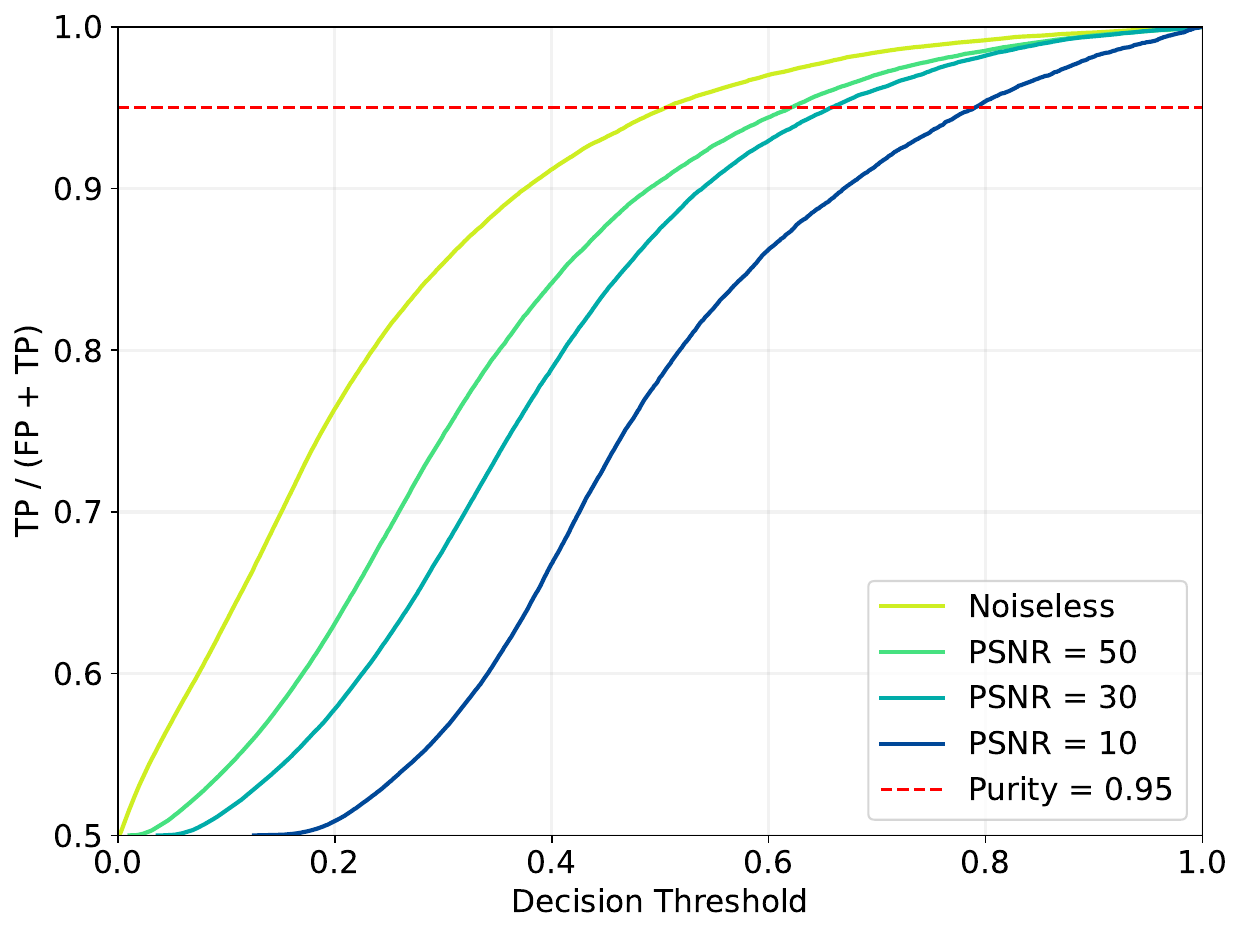}
\caption{The model's purity as a function of its decision threshold.}
\label{fig:Purity}
\hfill  % fill the gap between the two minipages % adjust the width of the minipage asded
\centering
\includegraphics[width=\linewidth]{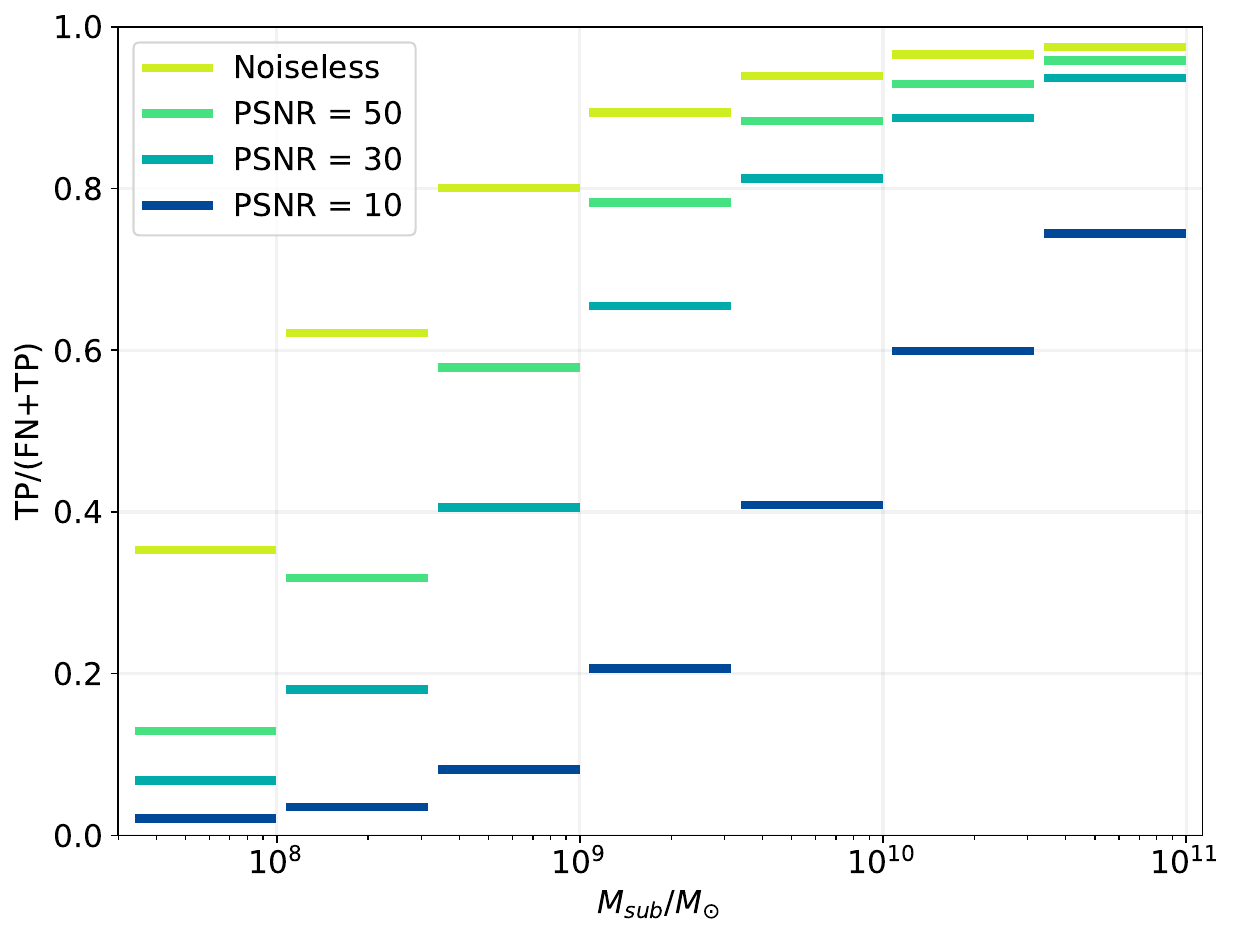}
\caption{Completeness of the model as a function of the decision threshold.}
\label{fig:Completeness}
\end{figure}

\begin{figure*}
    \centering
    \includegraphics[width=0.75\linewidth]{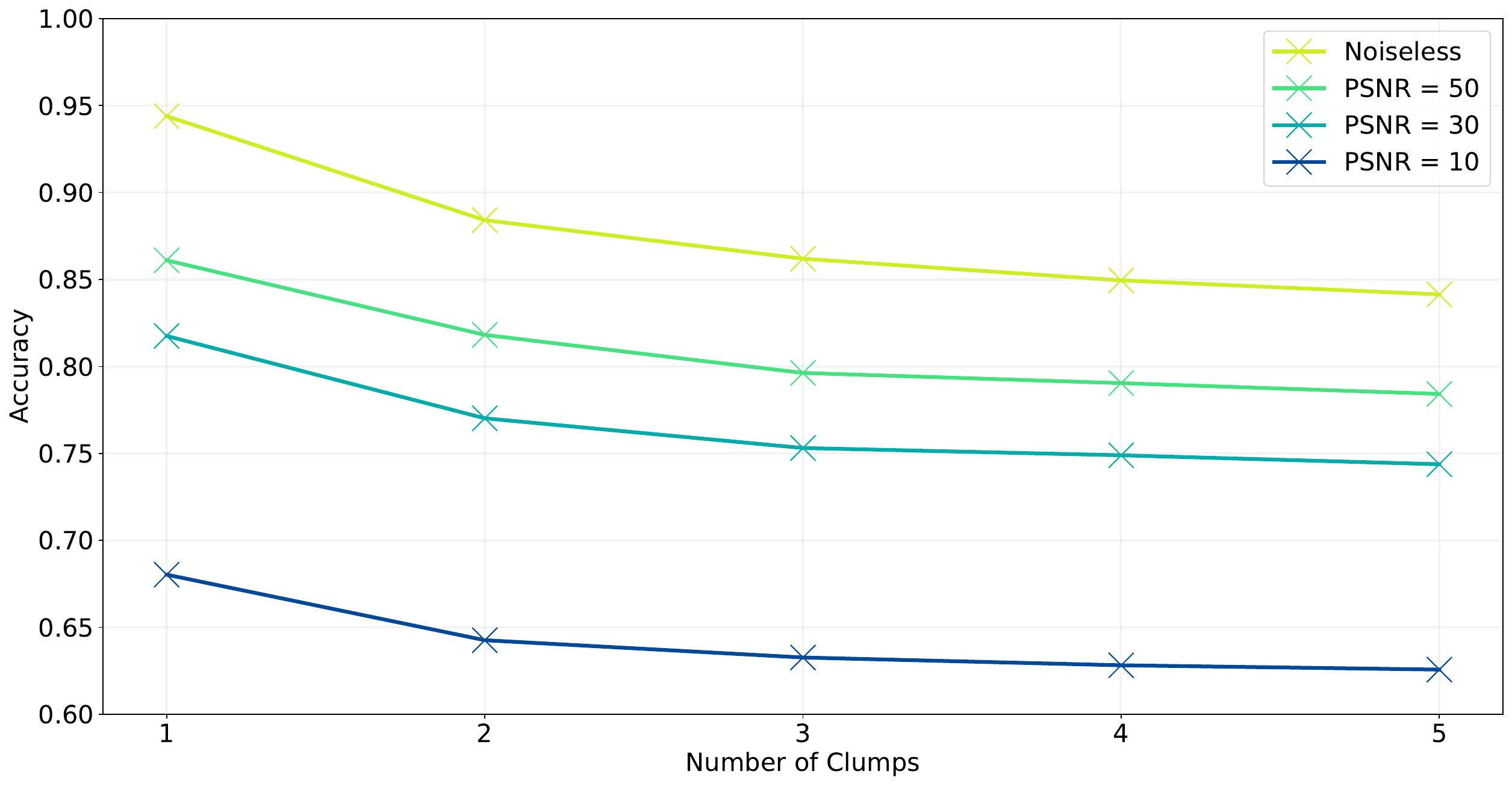}
    \caption{By dividing the test set into groups depending on the number of Sersic clumps in the source plane we can see how the model performances changes for more complex source structures.}
    \label{fig:acc_vs_clump}
\end{figure*}

Convolutional neural networks (CNN) have been very successful in computer vision related tasks such as image classification. Typically deeper CNN's are better for image classification as the later convolutional layers are able to learn higher order correlations between the basic features learnt in the earlier layers. However beyond a certain point the addition of extra layers has been empirically shown to reduce training accuracy \citep{7299173}. Mitigating this issue requires building the network architecture in such a way that it can bypass the extra convolutional layers if they don't further optimise the model during training. For this reason we use the ResNet50 architecture \citep{he_deep_2016} which implements this bypass functionality using residual skip connections.

The datasets used to train and test the model contained the full range of subhalo masses and source clump numbers. We found that the inclusion of higher mass subhalos meant the model was better able to learn how to identify lower mass subhalos improving both training speed and accuracy. Including examples of more obvious perturbations, such as when the subhalo mass is high and the source complexity is low, meant that the model had an easier time finding those same features when they're more subtle. 

We train the model in Tensorflow \citep{tensorflow2015-whitepaper} using the Adam optimisation algorithm with the default initial learning rate of 0.001 and sparse categorical cross-entropy to evaluate the loss. We simulate 800,000 lenses with a balanced number of perturbed and unperturbed images and split the dataset 80\% training/ 20\% test. The model was trained over 50 epochs, enough for the test accuracy to converge and the weights were taken from the epoch with the lowest test loss as this is when generalisability is maximised. We then repeat the process training on datasets with each combination of noise and image resolution. 

\hypertarget{Results-and-discussion}{%
\section{Results and Discussion}\label{Results}} 

We find that ResNet50 is able to identify subhalos down to $10^{7.5}M_{\odot}$ at noise levels with a PSNR of 30 and above, even with complex source morphologies. Here we delve into the confidence that we can have in these detections. The gold standard for any machine learning model is to attain both 100\% purity and completeness, but in practice some measure of trade off between the two is required and one needs to be prioritised over the other. Because neural networks operate as black boxes we want a high level of trust in any positive substructure detection and so minimising the false positive rate is important hence purity takes priority. For the scope of this project we set the requirement that the decision threshold must yield a purity of 95\%, that is to say that if the model were to make a detection there would be a 95\% probability that the detection is true. We chose this arbitrarily such that we maintain a high true positive rate without dramatically dropping the overall number of positive detections. 

We can alter the purity of the output by changing the decision threshold. Typically for a binary classification model the default is set to 0.5 but we can maximise purity or completeness by increasing or decreasing the threshold respectively. Figure \ref{fig:ROC} shows the receiver operating curve (ROC) for ResNet50 on the test set for an array of different noise levels. The open circle indicates the position on the curve corresponding to a decision threshold of 0.5. It's clear that as the data gets noisier the number of false positives increases, which isn't unexpected given noise can be mistaken as substructure by the model. What it does show is that to maintain our 95\% purity criteria we will need to raise the decision threshold for data with a lower PSNR. It's also worth noting that the overall performance of the model, dictated by the area under the curve (AUC), converges as the PSNR increases and you should be able to get close to noiseless performance on datasets with a PSNR greater than 50.

Figure \ref{fig:Purity} depicts the purity of the model's output as a function of the decision threshold. At a threshold of 0.5 the amount of noise can reduce the purity by up to 17\%. With that in mind we use the decision thresholds outlined in table \ref{tab:decision_thresholds}. The overall completeness for the corresponding decision thresholds are also listed, however this completeness is not uniform across subhalo masses. From figure \ref{fig:Completeness} we can see that the model struggles to capture most of the subhalos below a mass of $10^8M_{\odot}$ regardless of noise level. This has implications when attempting to reconstruct the subhalo mass function. The model exhibits a bias towards capturing more of the higher mass subhalo which could lead to an artificial suppression of the subhalo mass function in the low mass region. A full understanding of how a networks completeness changes as a function of subhalo mass is then necessary so that any resulting subhalo mass function can be scaled accordingly to avoid making any inaccurate conclusions about the nature of dark matter.

\begin{table}
    \centering
    \begin{tabular}{ c|c c }
         \hline
         PSNR  & Decision Threshold & Completeness \\
         \hline
         Noiseless & 0.51 & 0.79 \\
         50 & 0.62 & 0.64 \\
         30 & 0.66 & 0.56 \\
         10 & 0.79 & 0.29 \\
         \hline
    \end{tabular}
    \caption{Decision thresholds used for each noise level using ResNet50 which return a purity of 95\% and the overall completeness across the test set as a result.}
    \label{tab:decision_thresholds}
\end{table}

\begin{figure*}
    \centering
    \includegraphics[width=0.8\linewidth]{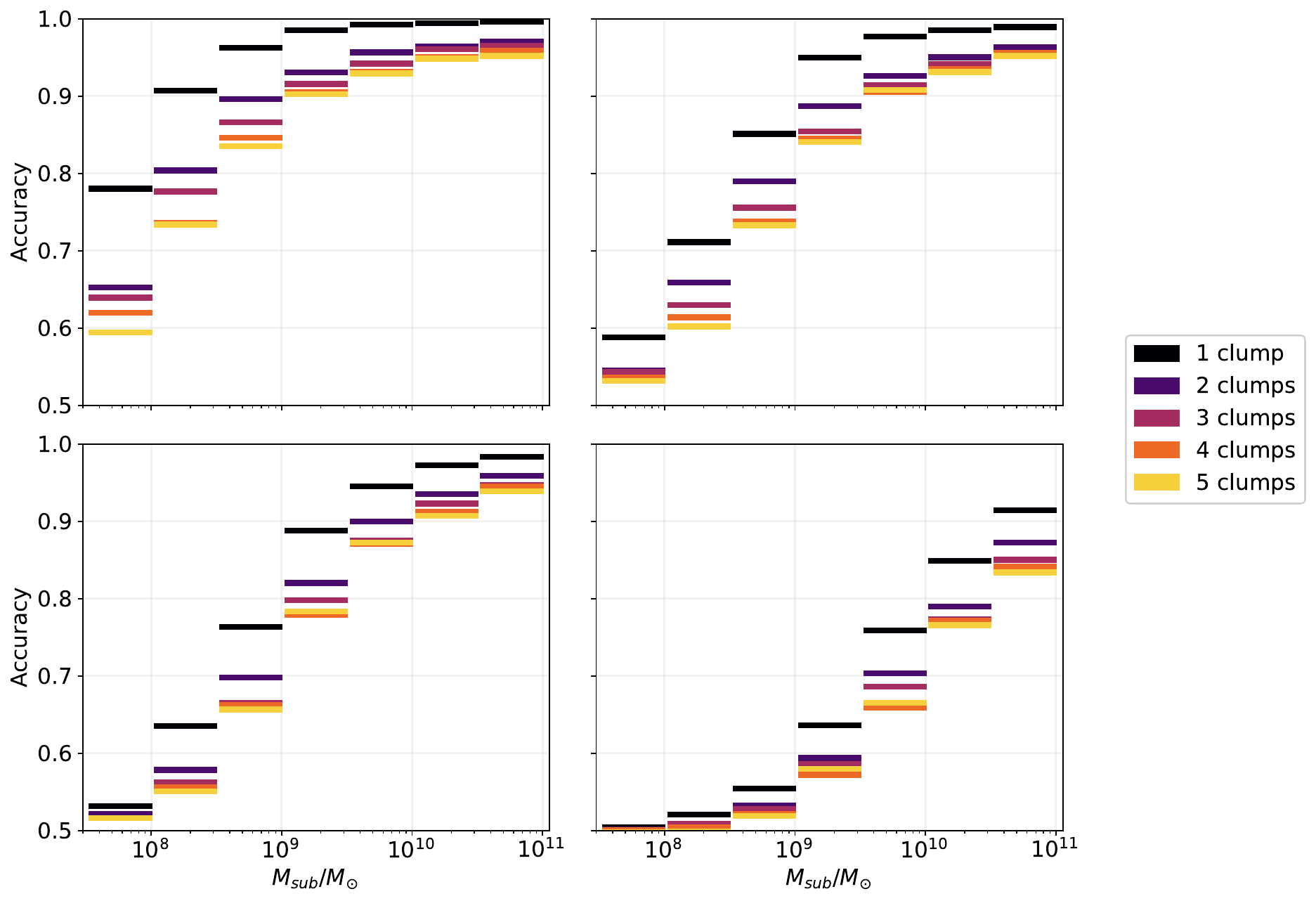}
    \caption{Model accuracy on the test set binned by both subhalo mass and number of Sersic clumps in the source plane. Each subhalo bin is populated with an equal number of randomly sampled unperturbed lenses from the test set to ensure each bin is balanced.}
    \label{fig:acc_v_submass}
\end{figure*}
\begin{figure*}
    \centering
    \includegraphics[width=0.85\linewidth]{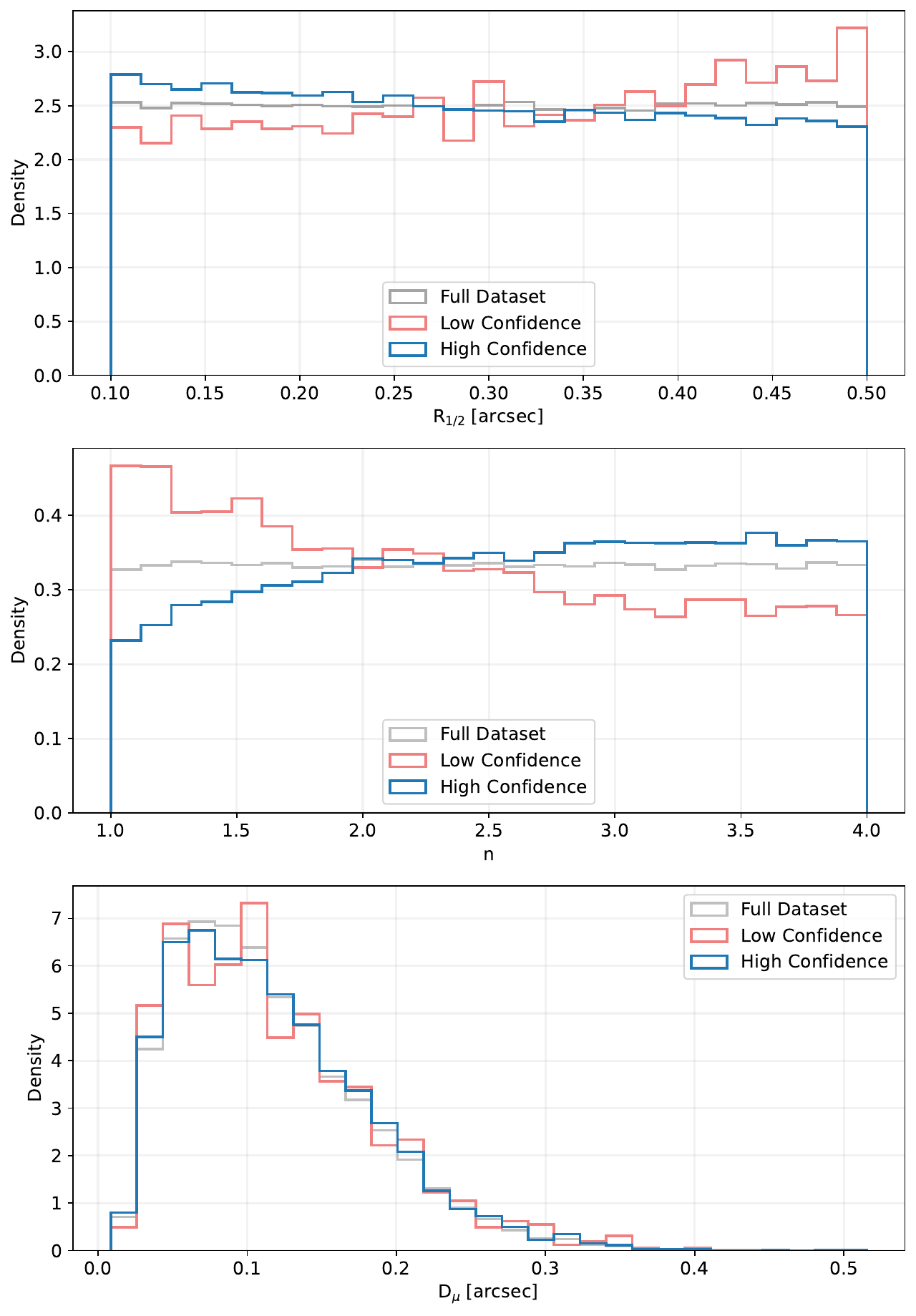}
    \caption{Distributions of the Sersic index, half-light radius, and average clump separation in images that the model classifies with high-confidence against images classified with low-confidence. We define high-confidence when the model provides a predicted output greater than 0.9, and low-confidence a predicted output between 0.45 and 0.55. The grey line shows the true distribution of the test set.}
    \label{fig:source_compactness_data}
\end{figure*}
\subsection{Model dependence on source complexity}
Here we analyse how the complexity of the source restricts the model's aptitude for successfully classifying images as either perturbed or unperturbed. Figure \ref{fig:acc_vs_clump} depicts the model's accuracy (the fraction of images correctly classified) as a function of the number of clumps in the source plane. From single-clump sources to 5-clump sources we see a 10\% drop in performance for the noiseless test set, and an 8\%, 7\%, and 5\% drop for images with a PSNR of 50, 30, and 10 respectively. This indicates that source complexity becomes less of an issue for noisier data as the noise becomes the dominant factor impeding the subhalo signal. Interestingly, complexity seems to have a diminishing effect on the model, with accuracy converging at a higher number of clumps. The implication from these results are that a subhalo is maximally detectable by a neural network when the source morphology is simple such as in an elliptical galaxy, but that even for more complex structures the model is still able to identify the perturbation. 

%% NOTE: This is for single subhalo, its not clear how much use the counter image is if there is a second perturber in the counter image.

 For the purpose of substructure detection, it is crucial that these models sufficiently probe the low-mass segment of the subhalo mass function and so we need to ensure that source complexity doesn't prevent the CNN from doing so. Figure \ref{fig:acc_v_submass} shows the model accuracy on the test sets binned by both subhalo mass and number of clumps in the source plane. Each mass bin is populated with a sample of unperturbed images from the test set equal to the number of perturbed images to ensure a balanced set. Looking at the noiseless example will give us insight into the inherent effects of source complexity, which has a larger effect on lower mass subhalos. We see a drop of almost 20\%  accuracy between single clump and 5 clump sources in the $10^{7.5}M_{\odot}$ to $10^{8}M_{\odot}$ mass range, compared to only a 5\% drop in the $10^{10.5}M_{\odot}$ to $10^{11}M_{\odot}$ range. Considering both the maximum magnification and deflection angle of a larger mass perturber is much higher than that of a lower mass perturber this is to be expected as the model is going to have a more difficult time finding weaker lensing signals and anything making that harder should have a more prominent effect. 
 
 We can see just how important high PSNR data is once we factor in the effects of noise. In the low mass bins we find a larger drop in performance across the single clump sources than the 5 clump sources with decreasing PSNR indicating that noise has a larger effect on lenses with simple sources. In the lowest mass bin and highest clump number, where the model has the most difficulty, we see a drop from 59\% in the noiseless data to 54\%, 51\% , and 50\% for a PSNR of 50, 30, and 10 respectively. Only in the lowest signal-to-noise data does the model fail to find any subhalos in the lowest mass bin. ResNet50 is therefore sophisticated enough to identify the lensing signal of a subhalo in the mass range low enough to significantly distinguish between CDM and other dark matter regimes even in the presence of complex source morphologies, however high signal data is necessary to maximise our chances of finding subhalos in this mass region. 

 \begin{figure*}
    \centering
    \includegraphics[width=0.7\linewidth]{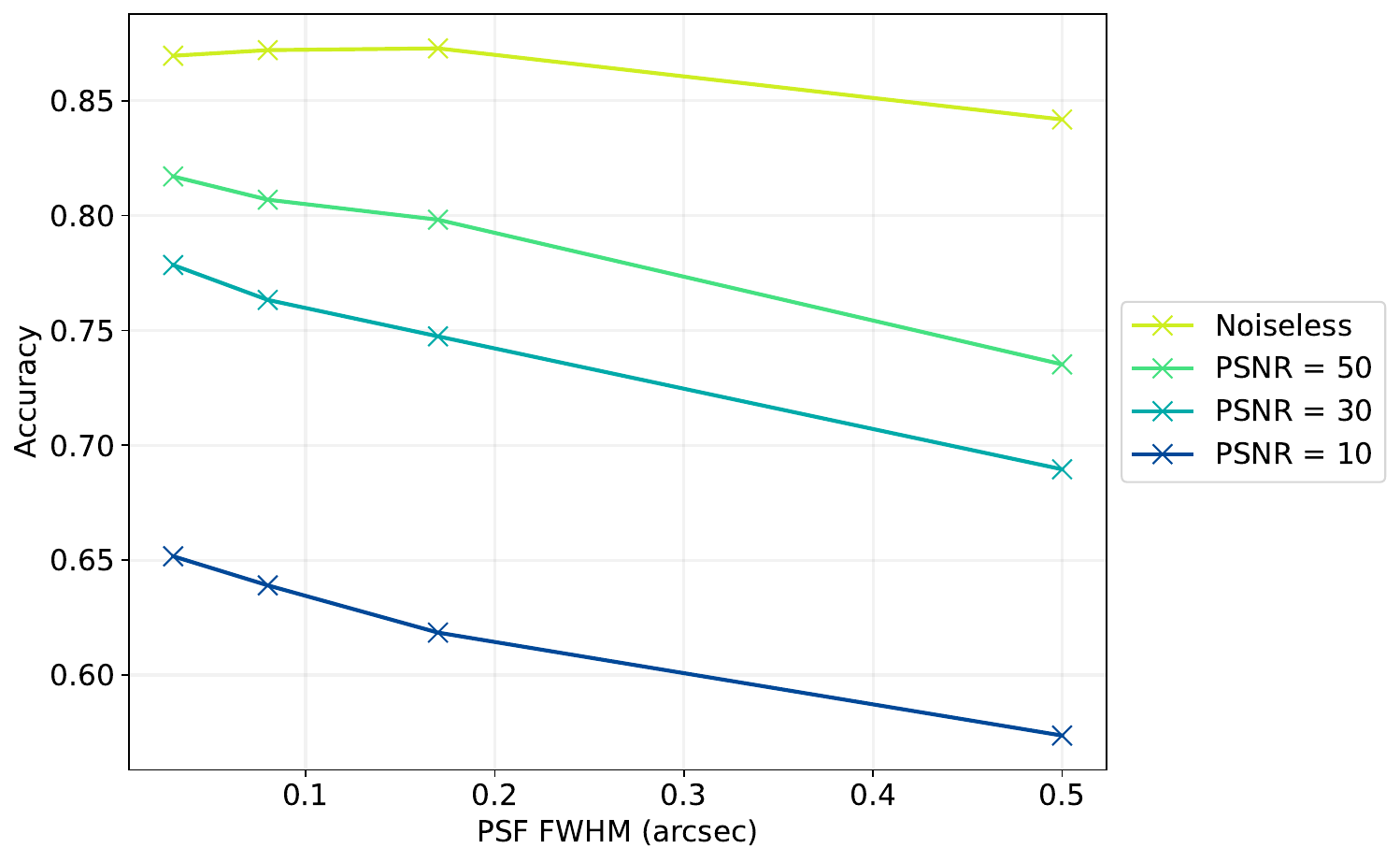}
    \caption{The accuracy of ResNet50 when classifying the test set as a function of the FWHM of the PSF. PSNR here is defined as the PSNR in the dataset with FWHM of 0.08"}
    \label{fig:acc_v_res}
\end{figure*}

\subsection{Model dependence on source compactness}

\begin{figure}
    \centering
    \includegraphics[width=0.9\linewidth]{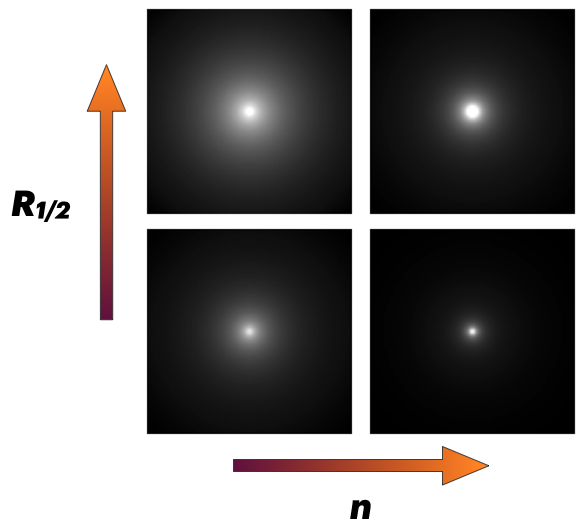}
    \caption{Diagram showing how $R_{1/2}$ and $n$ change the shape of the source light profile for a single clump.}
    \label{fig:source_shape}
\end{figure}

Complexity is just one way to parameterise the morphology; we can also look at the shape and distribution of the clumps. In this study we use circularly symmetric rather than elliptical Sersic profiles to somewhat simplify the analysis. In this case the shape of the individual clumps can be characterised by the Sersic index $n$ and the half-light radius $R_{1/2}$. Both variables dictate how diffuse or concentrated the clump is around its centroid, a visual example is given in figure \ref{fig:source_shape}. For a given source each clump is set to have the same value for both $n$ and $R_{1/2}$. To measure the distribution of the clumps in a source we take mean distance of each clump from the central clump. 
$$D_{\mu} = \sum_{n=2}^N\sqrt{(x_1^2-x_n^2)+(y_1^2-y_n^2)}$$
where N is the total number of clumps. More concentrated distributions will have lower mean distances. We can then analyse how these parameters impact the model's confidence in its predictions. We take all predictions across the test set and classify them as either high-confidence, where the output prediction is above 0.90, or low-confidence, where the output is between 0.45 and 0.55, and disregard the rest. We then compare the distribution of the above mentioned parameters which can be seen in figure \ref{fig:source_compactness_data}.

The bottom plot only considers images with greater than a single clump as $D_{\mu }$
can only be defined in these systems. Rather than doing a full noise analysis the results were taken from the predictions on data with a PSNR of 50 so they can be better generalised to real world data. 

Lenses with low-confidence predictions tend to have slightly larger $R_{1/2}$ and smaller $n$ while the opposite is true for high-confidence predictions meaning the model has an easier time identifying subhalos when the independent clumps are compact. Model confidence has a higher sensitivity to $n$ than $R_{1/2}$, likely a result of the different ways in which they influence the shape of the light profile. Where $n$ alters how quickly the surface brightness falls off with distance, $R_{1/2}$ effectively acts as a scaling parameter. Given each image is mean normalised before being fed into the network, this has less of an impact on the morphology of the images seen during training. It's important to note, even though the high confidence distributions skew from that of the full dataset the model still manages to recover much of the distribution, working well even on diffuse sources. This is likely a testament to the high dimensionality of the datasets parameter space. Intrestingly, we see no dependence with the models confidence on the distribution of clumps in the source. For a given image position, the source plane position that it traces back to changes only slightly when you add a subhalo perturbation; for a diffuse source the difference in surface brightness between these two points in the source plane is less than if the source were compact, meaning there is a smaller difference between the perturbed and unperturbed lens, making the subhalo harder to detect. However, a compact source yields a smaller lensed image, and hence the region probed for substructure in the lens is much smaller, reducing the chance of a subhalo perturbation being present. This means that if you aim to use only simple compact sources to maximise the lensing signal of a subhalo then you'll need a larger sample size before you'll find a perturbed lens. It may therefore be ideal to use lenses where the source is large, but has a lot of complex compact structure which, as made evident in the previous section, does not render the network unable to identify low mass subhalos.

\hypertarget{resolution}{%
\subsection{Detector resolution effects}}

\begin{figure*}
    \centering
    \includegraphics[width=\linewidth]{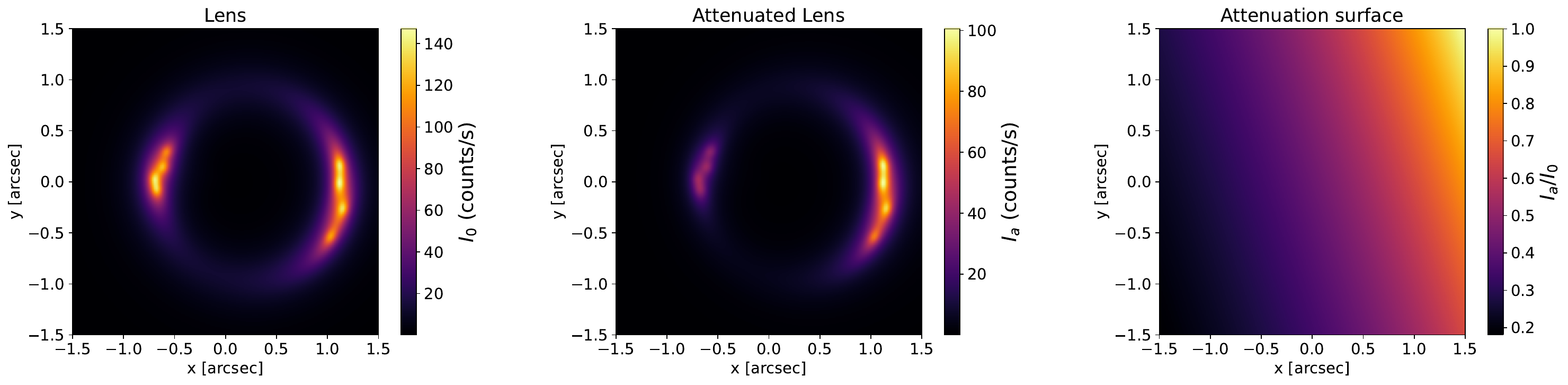}
    \caption{An example of a lens both before (left) and after (centre) applying the attenuation gradient (right).}
    \label{fig:attenuation_grad}
\end{figure*}
\begin{figure}
    \centering
    \includegraphics[width=\linewidth]{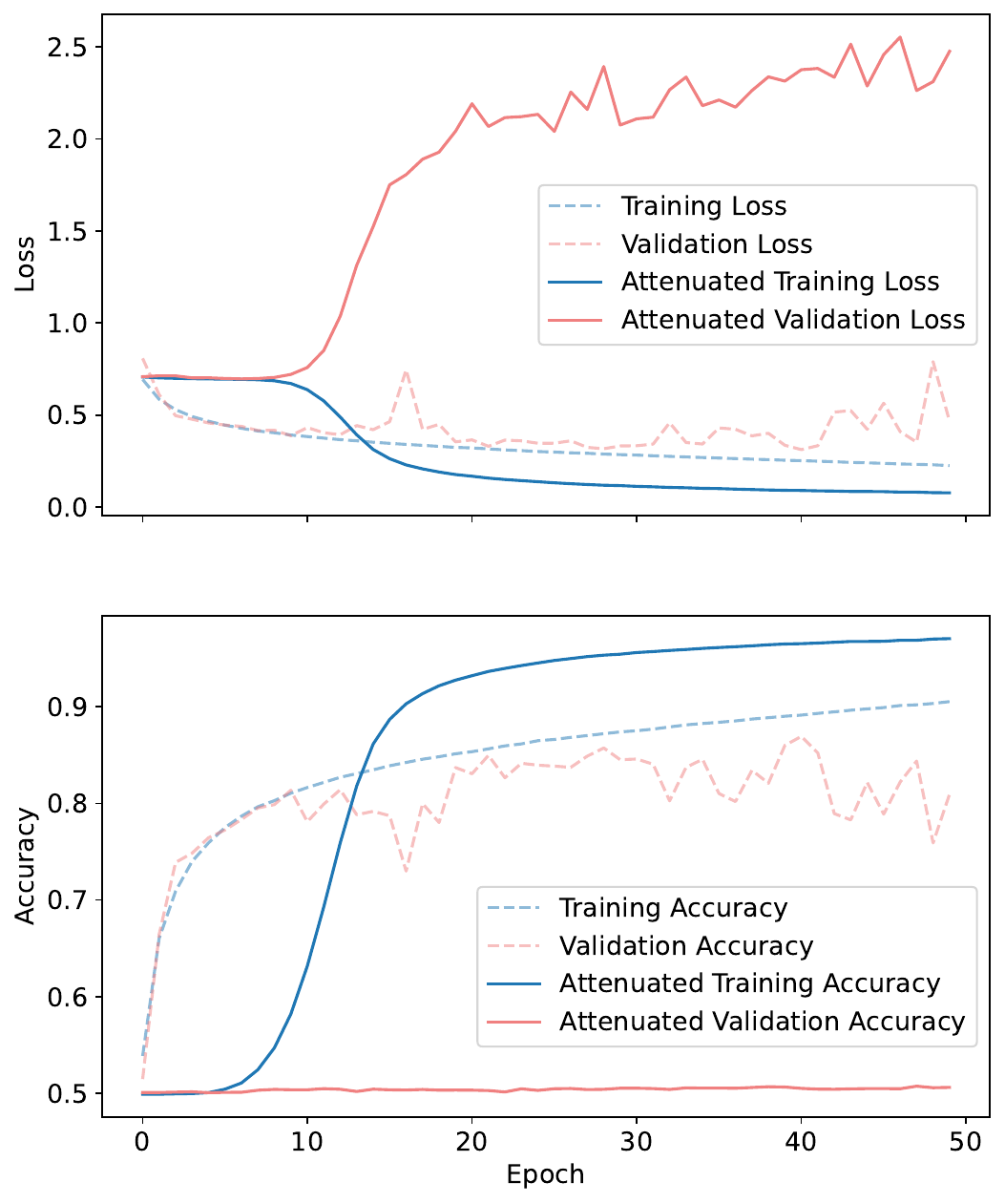}
    \caption{Training and validation loss (top) and accuracy (bottom) during training. Dashed lines show the typical learning trend and solid lines show trends when the lenses are artificially attenuated to change the flux ratios.}
    \label{fig:attn_train}
\end{figure}

With JWST data becoming increasingly available we expect high resolution followup of many of the galaxy-galaxy lenses found by surveys like DES \citep{2018ApJS..239...18A}. This improvement could have substantial benefits for substructure detection, though this new advantage has yet to be quantified. The expectation was that higher resolution data would be able to resolve the small scale deflections of the subhalos far better than lower resolution data. From figure \ref{fig:acc_v_res} we find that this statement, although true, is not as strong as expected. Between the JWST-like resolution and natural seeing we see around a 15\% drop in model accuracy regardless of noise, with exception to the noiseless data. Instead we see a drop of only a few percent, and practically no change in performance between the three highest resolved datasets. The general impression from these results is that while there is a benefit to highly resolved data, it doesn't seem as though the small scale structural perturbations to the image are the dominant feature being used by the model to identify subhalos. 

Subhalos perturb the structure of lensed images, but they also alter the flux. Some of the first evidence for subhalos were obtained by measuring the flux ratios between the multiple images in the lens. If the model is predominantly using these flux ratios then the spacial resolution of the image will be of little consequence. We can test this by artificially changing the flux ratios and retraining the model. In each of the resolution/noise combinations we apply an artificial linear attenuation gradient over each of the lenses using the following 
function:
$$I_a = I_0e^{ax+by}$$
where $I_a$ is the attentuated flux at position (x,y) in the image, $I_0$ is the unattenuated flux, and a and b are coefficients that define the steepness of the gradient and are taken from uniform distributions between $-3>a,b>3$. The freedom to either be negative or positive means the angle at which the gradient passes over the image covers a full 360$^{\circ}$. Attenuation is applied before the dataset is mean normalised. Figure \ref{fig:attenuation_grad} gives an example of an attenuated lens.

\begin{figure*}
    \centering
    \includegraphics[width=0.8\linewidth]{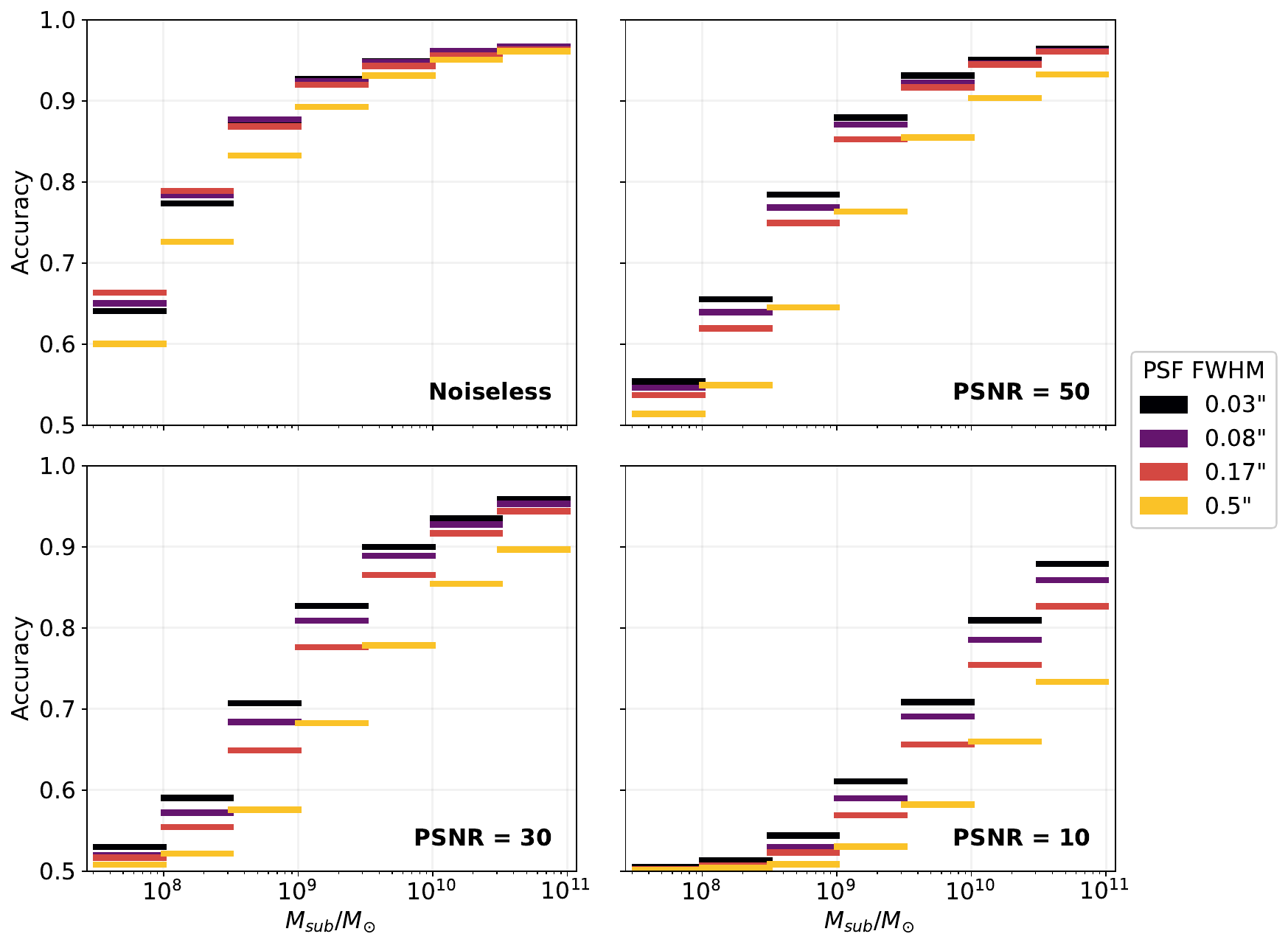}
    \caption{The accuracy of ResNet50 accross the full subhalo mass range in the test set for each of the different resolutions and noise levels.}
    \label{fig:acc_submass_res}
\end{figure*}

The expectation was that altering the flux ratios would have a some impact on model accuracy, but when training the model on the attenuated datasets it was unable to learn at any point across the allotted 50 epochs for any of the noise/resolution combinations and after 5-10 epoch that network began over-fitting to the training sets as seen in figure \ref{fig:attn_train}. This strongly suggests that not only is the flux ratio an important feature, but that it may be one of the first features the model uses to identify subhalos. From figure \ref{fig:acc_v_res} it's clear that the small scale features in the higher resolved images are learnt by the model, but without being able to learn from the flux ratios the network can't take advantage of the extra information. 

Though resolution doesn't seem to have as dramatic of an impact on the networks overall accuracy as expected, it's important to understand how exactly the performance changes on the lower mass subhalos given this is the region where we will be able to differentiate between different dark matter models. From figure \ref{fig:acc_submass_res} we make the following observations. In the noiseless regime resolution has very little effect on performance across all subhalo masses with the exception of natural seeing in which the majority of the drop in accuracy is seen in the low mass subhalos. Given the flux ratios do not change between datasets this shows that there are some structural features that the model is using even to identify these low mass subahlos. When factoring in noise the resolution certainly becomes more important. We see that even with the highest PSNR the drop in accuracy between the most resolved and least resolved dataset is of the order of 10\% in the low mass region, and has learned almost nothing on how to identify subhalos below $10^8 M_{\odot}$. With a PSNR of 30 we see a much bigger need to have higher resolved data, and with a PSNR of 10 at no resolution is the data sufficient to probe for the low mass subhalos. Given an infinite signal to noise ratio resolution has little impact on the detectability of subhalos below 0.17", meaning that even with the best quality data instruments limited by natural seeing are never an optimal choice for substructure detection. What becomes apparent once we add noise is that you need a higher signal to noise ratio for lower resolved data to get the same performance as on higher resolved data. For that reason, it is always preferable to aim for higher resolution data as for the same exposure time you will always have a better chance at detecting low mass subhalos. 

It's important to keep in mind that these datasets all use the same pixel scale, and it's not clear what impact this would have on performance. You would not expect data with a PSF FWHM of 0.5" to have such a fine pixel scale. Given resolution does have some impact on peformance, spatially resolved features are important for classification, information that would be lost to a larger pixel scale. And so realistically we would not expect even data with a FWHM of 0.17" to perform this well. The pixel scale is consistent with drizzled HST data and raw JWST data, so we don't expect performance to be that different than what could be expected realistically. With a resolution of 0.03" the PSF is not properly sampled and so pixel scale will have a limiting impact on performance. We might expect a model to have even better subhalo detectability on drizzled JWST data. 

\hypertarget{conclusion}{%
\section{Conclusion}\label{conclusion}}

In this work we analyse ResNet50's ability to identify the lensing signal of a single subhalo against the presence of increasingly more complex source morphologies and identify the features of sources that maximise detectability. 
To maintain a high level of purity it's necessary to increase the decision threshold, however this issue becomes less of a problem for images with a high PSNR above 50. Increasing purity does sacrifice completeness which is not uniform across subhalo mass introducing a selection bias towards high mass subhalos. Completeness of a network should be well understood so that any inferred subhalo mass functions could be scaled accordingly.

We found that the complexity of the source has a diminishing effect on the accuracy of ResNet50 and sources with 3 or more clumps yield similar performance. Complexity had the largest effect on low mass subhalos but as long as PSNR > 10 the model is still able to achieve an accuracy above random guessing showing that complexity does not completely extinguish the subhalo signal in the mass range needed to differentiate between dark matter regimes. 

The model was more confident in its predictions when the clumps in the source galaxy were compact, but was indifferent the their spatial distribution. The subhalo signal was strongest when there is just a single highly concentrated clump in the source. The trade-off here is that the region in the lens probed by the source light is much smaller and so the chances of it being perturbed are lower. It's therefore preferable to use lenses with larger images that posses a lot of fine compact structure (such as a face on spiral). Despite compact sources improving model confidence, it was still able to identify a large number of subhalos in lenses with diffuse sources. 

We conclude that complex morphologies do not prevent convolutional neural networks from identifying the perturbation from a single subhalo in strong galaxy-galaxy lenses even at masses low enough to distinguish between CDM and other dark matter models. This analysis serves as a test of a CNN's sensitivity to source morphology, in reality, simple binary classification is not sufficient to measure the subhalo mass function. The model would be improved by returning the mass of the subhalo to make any claim about dark matter. If one wants to constrain the full subhalo mass function then it's important to note that these models are biased towards detecting higher mass subhalos and subhalo counts per mass range should be scaled to account for this. With a high enough signal-to-noise ratio source complexity can have some additional impact on these counts which will also need to be accounted for.

Reducing the FWHM of the dataset's PSF increases the ResNet's ability to detect subhalos, though the effect is not as strong as was expected. This is largely to do with the network using flux ratio anomalies, which are conserved across different resolutions, to look for perturbations. We find it is always best to aim for highly resolved photometery as there is certainly additional information in the resolved small scale structure that can be learned by the model when looking for low mass subhalos and would require a lower exposure time to yield the same performance than lower resolved data.

We do highlight that this analysis is used more to establish some fundamental relationships between the detectability of a subhalo to a CNN and the source morphology, but the results can't yet be generalised to real-world photometric data. In reality there are many systematics that need to be considered before these models can be confidently applied to real lenses. In particular, things like full subhalo populations, line-of-sight halos,  and anisotropy in the main lens density profile are all features in real galaxies that could be mistaken for a subhalo. We also ignore any affect that baryons may have on the density profile of the subhalo. Though at masses below $10^8M_{\odot}$ the baryon fraction is small enough for the subhalo to be considered "dark", we expect this to be more of an issue for larger subhalos. The parametric model used for the source surface brightness is only analogous to the types of morphological features you would expect in $1<\text{z}<2$ galaxies, but is not fully representative. We intend to account for much of this in future work, using more realistic source galaxies potentially from simulations, and building in as many systematics and degeneracies such as those mentioned above on top of others such as correlated noise that you might find in drizzled data. 

\hypertarget{Acknowledgements}{
\section*{Acknowledgements}\label{acknowledgements}
TJH would like to acknowledge support from the Australian Research Council Centre of Excellence for Dark Matter Particle Physics (CDM, CE200100008). This work was performed on the OzSTAR and Ngarrgu Tinderbeek facilities at Swinburne University of Technology.
}

\hypertarget{Data Availability}{%
\section*{Data Availability}\label{data availability}}
The data underlying this article will be shared on reasonable request to the corresponding author.



\bibliographystyle{mnras}
\bibliography{main} % if your bibtex file is called example.bib

\begin{thebibliography}{}
\makeatletter
\relax
\def\mn@urlcharsother{\let\do\@makeother \do\$\do\&\do\#\do\^\do\_\do\%\do\~}
\def\mn@doi{\begingroup\mn@urlcharsother \@ifnextchar [ {\mn@doi@}
  {\mn@doi@[]}}
\def\mn@doi@[#1]#2{\def\@tempa{#1}\ifx\@tempa\@empty \href
  {http://dx.doi.org/#2} {doi:#2}\else \href {http://dx.doi.org/#2} {#1}\fi
  \endgroup}
\def\mn@eprint#1#2{\mn@eprint@#1:#2::\@nil}
\def\mn@eprint@arXiv#1{\href {http://arxiv.org/abs/#1} {{\tt arXiv:#1}}}
\def\mn@eprint@dblp#1{\href {http://dblp.uni-trier.de/rec/bibtex/#1.xml}
  {dblp:#1}}
\def\mn@eprint@#1:#2:#3:#4\@nil{\def\@tempa {#1}\def\@tempb {#2}\def\@tempc
  {#3}\ifx \@tempc \@empty \let \@tempc \@tempb \let \@tempb \@tempa \fi \ifx
  \@tempb \@empty \def\@tempb {arXiv}\fi \@ifundefined
  {mn@eprint@\@tempb}{\@tempb:\@tempc}{\expandafter \expandafter \csname
  mn@eprint@\@tempb\endcsname \expandafter{\@tempc}}}

\bibitem[\protect\citeauthoryear{Abadi et~al.,}{Abadi
  et~al.}{2015}]{tensorflow2015-whitepaper}
Abadi M.,  et~al., 2015, {TensorFlow}: Large-Scale Machine Learning on
  Heterogeneous Systems, \url {https://www.tensorflow.org/}

\bibitem[\protect\citeauthoryear{{Abbott} et~al.,}{{Abbott}
  et~al.}{2018}]{2018ApJS..239...18A}
{Abbott} T.~M.~C.,  et~al., 2018, \mn@doi [\apjs] {10.3847/1538-4365/aae9f0},
  \href {https://ui.adsabs.harvard.edu/abs/2018ApJS..239...18A} {239, 18}

\bibitem[\protect\citeauthoryear{{Baltz}, {Marshall}  \& {Oguri}}{{Baltz}
  et~al.}{2009}]{2009JCAP...01..015B}
{Baltz} E.~A.,  {Marshall} P.,   {Oguri} M.,  2009, \mn@doi [\jcap]
  {10.1088/1475-7516/2009/01/015}, \href
  {https://ui.adsabs.harvard.edu/abs/2009JCAP...01..015B} {2009, 015}

\bibitem[\protect\citeauthoryear{Birrer et~al.,}{Birrer
  et~al.}{2021}]{Birrer2021}
Birrer S.,  et~al., 2021, \mn@doi [Journal of Open Source Software]
  {10.21105/joss.03283}, 6, 3283

\bibitem[\protect\citeauthoryear{{Boylan-Kolchin}, {Bullock}  \&
  {Kaplinghat}}{{Boylan-Kolchin} et~al.}{2011}]{2011MNRAS.415L..40B}
{Boylan-Kolchin} M.,  {Bullock} J.~S.,   {Kaplinghat} M.,  2011, \mn@doi
  [\mnras] {10.1111/j.1745-3933.2011.01074.x}, \href
  {https://ui.adsabs.harvard.edu/abs/2011MNRAS.415L..40B} {415, L40}

\bibitem[\protect\citeauthoryear{Dalal \& Kochanek}{Dalal \&
  Kochanek}{2002}]{Dalal_2002}
Dalal N.,  Kochanek C.~S.,  2002, \mn@doi [The Astrophysical Journal]
  {10.1086/340303}, 572, 25

\bibitem[\protect\citeauthoryear{{Davis}, {Efstathiou}, {Frenk}  \&
  {White}}{{Davis} et~al.}{1985}]{1985ApJ...292..371D}
{Davis} M.,  {Efstathiou} G.,  {Frenk} C.~S.,   {White} S.~D.~M.,  1985,
  \mn@doi [\apj] {10.1086/163168}, \href
  {https://ui.adsabs.harvard.edu/abs/1985ApJ...292..371D} {292, 371}

\bibitem[\protect\citeauthoryear{Davis, Huterer  \& Krauss}{Davis
  et~al.}{2003}]{Davis_Huterer_Krauss_2003}
Davis A.~N.,  Huterer D.,   Krauss L.~M.,  2003, \mn@doi [Monthly Notices of
  the Royal Astronomical Society] {10.1046/j.1365-8711.2003.06789.x}, 344,
  1029–1040

\bibitem[\protect\citeauthoryear{Despali, Vegetti, White, Giocoli  \& van~den
  Bosch}{Despali et~al.}{2018}]{10.1093/mnras/sty159}
Despali G.,  Vegetti S.,  White S. D.~M.,  Giocoli C.,   van~den Bosch F.~C.,
  2018, \mn@doi [Monthly Notices of the Royal Astronomical Society]
  {10.1093/mnras/sty159}, 475, 5424

\bibitem[\protect\citeauthoryear{Diaz~Rivero \& Dvorkin}{Diaz~Rivero \&
  Dvorkin}{2020}]{diaz_rivero_direct_2020}
Diaz~Rivero A.,  Dvorkin C.,  2020, \mn@doi [Phys. Rev. D]
  {10.1103/PhysRevD.101.023515}, 101, 023515

\bibitem[\protect\citeauthoryear{{Diemer} \& {Joyce}}{{Diemer} \&
  {Joyce}}{2019}]{2019ApJ...871..168D}
{Diemer} B.,  {Joyce} M.,  2019, \mn@doi [\apj] {10.3847/1538-4357/aafad6},
  \href {https://ui.adsabs.harvard.edu/abs/2019ApJ...871..168D} {871, 168}

\bibitem[\protect\citeauthoryear{{Frenk} \& {White}}{{Frenk} \&
  {White}}{2012}]{2012AnP...524..507F}
{Frenk} C.~S.,  {White} S.~D.~M.,  2012, \mn@doi [Annalen der Physik]
  {10.1002/andp.201200212}, \href
  {https://ui.adsabs.harvard.edu/abs/2012AnP...524..507F} {524, 507}

\bibitem[\protect\citeauthoryear{{Fruchter} \& {Hook}}{{Fruchter} \&
  {Hook}}{2002}]{2002PASP..114..144F}
{Fruchter} A.~S.,  {Hook} R.~N.,  2002, \mn@doi [\pasp] {10.1086/338393}, \href
  {https://ui.adsabs.harvard.edu/abs/2002PASP..114..144F} {114, 144}

\bibitem[\protect\citeauthoryear{Gilman}{Gilman}{2023}]{PyHalo}
Gilman D.,  2023, PyHalo, \url{https://github.com/dangilman/pyHalo}

\bibitem[\protect\citeauthoryear{{Gondolo} \& {Gelmini}}{{Gondolo} \&
  {Gelmini}}{1991}]{1991NuPhB.360..145G}
{Gondolo} P.,  {Gelmini} G.,  1991, \mn@doi [Nuclear Physics B]
  {10.1016/0550-3213(91)90438-4}, \href
  {https://ui.adsabs.harvard.edu/abs/1991NuPhB.360..145G} {360, 145}

\bibitem[\protect\citeauthoryear{Guan-wen, Ze-sen  \& Xu}{Guan-wen
  et~al.}{2018}]{GUANWEN201820}
Guan-wen F.,  Ze-sen L.,   Xu K.,  2018, \mn@doi [Chinese Astronomy and
  Astrophysics] {https://doi.org/10.1016/j.chinastron.2018.01.002}, 42, 20

\bibitem[\protect\citeauthoryear{He \& Sun}{He \& Sun}{2015}]{7299173}
He K.,  Sun J.,  2015, in 2015 IEEE Conference on Computer Vision and Pattern
  Recognition (CVPR). IEEE Computer Society, Los Alamitos, CA, USA, pp
  5353--5360, \mn@doi{10.1109/CVPR.2015.7299173}, \url
  {https://doi.ieeecomputersociety.org/10.1109/CVPR.2015.7299173}

\bibitem[\protect\citeauthoryear{He, Zhang, Ren  \& Sun}{He
  et~al.}{2016}]{he_deep_2016}
He K.,  Zhang X.,  Ren S.,   Sun J.,  2016, in Deep Residual Learning for Image
  Recognition. pp 770--778, \url
  {https://openaccess.thecvf.com/content_cvpr_2016/html/He_Deep_Residual_Learning_CVPR_2016_paper.html}

\bibitem[\protect\citeauthoryear{Hu, Barkana  \& Gruzinov}{Hu
  et~al.}{2000}]{PhysRevLett.85.1158}
Hu W.,  Barkana R.,   Gruzinov A.,  2000, \mn@doi [Phys. Rev. Lett.]
  {10.1103/PhysRevLett.85.1158}, 85, 1158

\bibitem[\protect\citeauthoryear{{Ishiyama}}{{Ishiyama}}{2014}]{2014ApJ...788...27I}
{Ishiyama} T.,  2014, \mn@doi [\apj] {10.1088/0004-637X/788/1/27}, \href
  {https://ui.adsabs.harvard.edu/abs/2014ApJ...788...27I} {788, 27}

\bibitem[\protect\citeauthoryear{{Klypin}, {Kravtsov}, {Valenzuela}  \&
  {Prada}}{{Klypin} et~al.}{1999}]{1999ApJ...522...82K}
{Klypin} A.,  {Kravtsov} A.~V.,  {Valenzuela} O.,   {Prada} F.,  1999, \mn@doi
  [\apj] {10.1086/307643}, \href
  {https://ui.adsabs.harvard.edu/abs/1999ApJ...522...82K} {522, 82}

\bibitem[\protect\citeauthoryear{Kochanek \& Dalal}{Kochanek \&
  Dalal}{2004}]{Kochanek_2004}
Kochanek C.~S.,  Dalal N.,  2004, \mn@doi [The Astrophysical Journal]
  {10.1086/421436}, 610, 69

\bibitem[\protect\citeauthoryear{{Koekemoer} et~al.,}{{Koekemoer}
  et~al.}{2007}]{2007ApJS..172..196K}
{Koekemoer} A.~M.,  et~al., 2007, \mn@doi [\apjs] {10.1086/520086}, \href
  {https://ui.adsabs.harvard.edu/abs/2007ApJS..172..196K} {172, 196}

\bibitem[\protect\citeauthoryear{Koopmans}{Koopmans}{2005}]{10.1111/j.1365-2966.2005.09523.x}
Koopmans L. V.~E.,  2005, \mn@doi [Monthly Notices of the Royal Astronomical
  Society] {10.1111/j.1365-2966.2005.09523.x}, 363, 1136

\bibitem[\protect\citeauthoryear{{Li}, {Frenk}, {Cole}, {Wang}  \& {Gao}}{{Li}
  et~al.}{2017}]{2017MNRAS.468.1426L}
{Li} R.,  {Frenk} C.~S.,  {Cole} S.,  {Wang} Q.,   {Gao} L.,  2017, \mn@doi
  [\mnras] {10.1093/mnras/stx554}, \href
  {https://ui.adsabs.harvard.edu/abs/2017MNRAS.468.1426L} {468, 1426}

\bibitem[\protect\citeauthoryear{{Mao} \& {Schneider}}{{Mao} \&
  {Schneider}}{1998}]{1998MNRAS.295..587M}
{Mao} S.,  {Schneider} P.,  1998, \mn@doi [\mnras]
  {10.1046/j.1365-8711.1998.01319.x}, \href
  {https://ui.adsabs.harvard.edu/abs/1998MNRAS.295..587M} {295, 587}

\bibitem[\protect\citeauthoryear{{Mashchenko}, {Couchman}  \&
  {Wadsley}}{{Mashchenko} et~al.}{2006}]{2006Natur.442..539M}
{Mashchenko} S.,  {Couchman} H.~M.~P.,   {Wadsley} J.,  2006, \mn@doi [\nat]
  {10.1038/nature04944}, \href
  {https://ui.adsabs.harvard.edu/abs/2006Natur.442..539M} {442, 539}

\bibitem[\protect\citeauthoryear{{Meszaros}}{{Meszaros}}{1974}]{1974A&A....37..225M}
{Meszaros} P.,  1974, \aap, \href
  {https://ui.adsabs.harvard.edu/abs/1974A&A....37..225M} {37, 225}

\bibitem[\protect\citeauthoryear{More, McKean, More, Porcas, Koopmans  \&
  Garrett}{More et~al.}{2009}]{More_McKean_More_Porcas_Koopmans_Garrett_2009}
More A.,  McKean J.~P.,  More S.,  Porcas R.~W.,  Koopmans L. V.~E.,   Garrett
  M.~A.,  2009, \mn@doi [Monthly Notices of the Royal Astronomical Society]
  {10.1111/j.1365-2966.2008.14342.x}, 394, 174–190

\bibitem[\protect\citeauthoryear{Morgan, Nord, Birrer, Lin  \& Poh}{Morgan
  et~al.}{2021}]{deeplenstronomy}
Morgan R.,  Nord B.,  Birrer S.,  Lin J. Y.-Y.,   Poh J.,  2021, \mn@doi
  [Journal of Open Source Software] {10.21105/joss.02854}, 6, 2854

\bibitem[\protect\citeauthoryear{Navarro}{Navarro}{1996}]{navarro_structure_1996}
Navarro J.~F.,  1996, \mn@doi [Symposium - International Astronomical Union]
  {10.1017/S0074180900232452}, 171, 255

\bibitem[\protect\citeauthoryear{Ostdiek, Rivero  \& Dvorkin}{Ostdiek
  et~al.}{2022}]{ostdiek_extracting_2022}
Ostdiek B.,  Rivero A.~D.,   Dvorkin C.,  2022, \mn@doi [{ApJ}]
  {10.3847/1538-4357/ac2d8d}, 927, 83

\bibitem[\protect\citeauthoryear{{Sawala} et~al.,}{{Sawala}
  et~al.}{2016}]{2016MNRAS.456...85S}
{Sawala} T.,  et~al., 2016, \mn@doi [\mnras] {10.1093/mnras/stv2597}, \href
  {https://ui.adsabs.harvard.edu/abs/2016MNRAS.456...85S} {456, 85}

\bibitem[\protect\citeauthoryear{{Schneider}, {Smith}  \& {Reed}}{{Schneider}
  et~al.}{2013a}]{2013MNRAS.433.1573S}
{Schneider} A.,  {Smith} R.~E.,   {Reed} D.,  2013a, \mn@doi [\mnras]
  {10.1093/mnras/stt829}, \href
  {https://ui.adsabs.harvard.edu/abs/2013MNRAS.433.1573S} {433, 1573}

\bibitem[\protect\citeauthoryear{Schneider, Smith  \& Reed}{Schneider
  et~al.}{2013b}]{10.1093/mnras/stt829}
Schneider A.,  Smith R.~E.,   Reed D.,  2013b, \mn@doi [Monthly Notices of the
  Royal Astronomical Society] {10.1093/mnras/stt829}, 433, 1573

\bibitem[\protect\citeauthoryear{Sengül, Tsang, Diaz~Rivero, Dvorkin, Zhu  \&
  Seljak}{Sengül et~al.}{2020}]{PhysRevD.102.063502}
Sengül A.~C.,  Tsang A.,  Diaz~Rivero A.,  Dvorkin C.,  Zhu H.-M.,   Seljak
  U.,  2020, \mn@doi [Phys. Rev. D] {10.1103/PhysRevD.102.063502}, 102, 063502

\bibitem[\protect\citeauthoryear{Sengül, Dvorkin, Ostdiek  \& Tsang}{Sengül
  et~al.}{2022}]{sengul_substructure_2021}
Sengül A.~C.,  Dvorkin C.,  Ostdiek B.,   Tsang A.,  2022, \mn@doi [Monthly
  Notices of the Royal Astronomical Society] {10.1093/mnras/stac1967}, 515,
  4391

\bibitem[\protect\citeauthoryear{{Sersic}}{{Sersic}}{1968}]{1968adga.book.....S}
{Sersic} J.~L.,  1968, {Atlas de Galaxias Australes}.
Cordoba

\bibitem[\protect\citeauthoryear{{Springel} et~al.,}{{Springel}
  et~al.}{2005}]{2005Natur.435..629S}
{Springel} V.,  et~al., 2005, \mn@doi [\nat] {10.1038/nature03597}, \href
  {https://ui.adsabs.harvard.edu/abs/2005Natur.435..629S} {435, 629}

\bibitem[\protect\citeauthoryear{{Springel}, {Frenk}  \& {White}}{{Springel}
  et~al.}{2006}]{2006Natur.440.1137S}
{Springel} V.,  {Frenk} C.~S.,   {White} S. D.~M.,  2006, \mn@doi [\nat]
  {10.1038/nature04805}, \href
  {https://ui.adsabs.harvard.edu/abs/2006Natur.440.1137S} {440, 1137}

\bibitem[\protect\citeauthoryear{Tran et~al.,}{Tran et~al.}{2022}]{Tran_2022}
Tran K.-V.~H.,  et~al., 2022, \mn@doi [The Astronomical Journal]
  {10.3847/1538-3881/ac7da2}, 164, 148

\bibitem[\protect\citeauthoryear{Vegetti \& Koopmans}{Vegetti \&
  Koopmans}{2009}]{vegetti_bayesian_2009}
Vegetti S.,  Koopmans L. V.~E.,  2009, \mn@doi [Monthly Notices of the Royal
  Astronomical Society] {10.1111/j.1365-2966.2008.14005.x}, 392, 945

\bibitem[\protect\citeauthoryear{Vegetti, Koopmans, Bolton, Treu  \&
  Gavazzi}{Vegetti et~al.}{2010}]{vegetti_detection_2010}
Vegetti S.,  Koopmans L. V.~E.,  Bolton A.,  Treu T.,   Gavazzi R.,  2010,
  \mn@doi [Monthly Notices of the Royal Astronomical Society]
  {10.1111/j.1365-2966.2010.16865.x}, 408, 1969

\bibitem[\protect\citeauthoryear{Vegetti, Lagattuta, {McKean}, Auger, Fassnacht
   \& Koopmans}{Vegetti et~al.}{2012}]{vegetti_gravitational_2012}
Vegetti S.,  Lagattuta D.~J.,  {McKean} J.~P.,  Auger M.~W.,  Fassnacht C.~D.,
   Koopmans L. V.~E.,  2012, \mn@doi [Nature] {10.1038/nature10669}, 481, 341

\bibitem[\protect\citeauthoryear{Wagner-Carena, Park, Birrer, Marshall, Roodman
   \& Wechsler}{Wagner-Carena et~al.}{2021}]{wagner-carena_hierarchical_2021}
Wagner-Carena S.,  Park J.~W.,  Birrer S.,  Marshall P.~J.,  Roodman A.,
  Wechsler R.~H.,  2021, \mn@doi [The Astrophysical Journal]
  {10.3847/1538-4357/abdf59}, 909, 187

\bibitem[\protect\citeauthoryear{{Wagner-Carena}, {Aalbers}, {Birrer},
  {Nadler}, {Darragh-Ford}, {Marshall}  \& {Wechsler}}{{Wagner-Carena}
  et~al.}{2023}]{wagner-carena_images_2022}
{Wagner-Carena} S.,  {Aalbers} J.,  {Birrer} S.,  {Nadler} E.~O.,
  {Darragh-Ford} E.,  {Marshall} P.~J.,   {Wechsler} R.~H.,  2023, \mn@doi
  [\apj] {10.3847/1538-4357/aca525}, \href
  {https://ui.adsabs.harvard.edu/abs/2023ApJ...942...75W} {942, 75}

\bibitem[\protect\citeauthoryear{{Walker} \& {Pe{\~n}arrubia}}{{Walker} \&
  {Pe{\~n}arrubia}}{2011}]{2011ApJ...742...20W}
{Walker} M.~G.,  {Pe{\~n}arrubia} J.,  2011, \mn@doi [\apj]
  {10.1088/0004-637X/742/1/20}, \href
  {https://ui.adsabs.harvard.edu/abs/2011ApJ...742...20W} {742, 20}

\bibitem[\protect\citeauthoryear{{Weinberg} \& {Katz}}{{Weinberg} \&
  {Katz}}{2002}]{2002ApJ...580..627W}
{Weinberg} M.~D.,  {Katz} N.,  2002, \mn@doi [\apj] {10.1086/343847}, \href
  {https://ui.adsabs.harvard.edu/abs/2002ApJ...580..627W} {580, 627}

\bibitem[\protect\citeauthoryear{{White}, {Frenk}  \& {Davis}}{{White}
  et~al.}{1983}]{1983ApJ...274L...1W}
{White} S.~D.~M.,  {Frenk} C.~S.,   {Davis} M.,  1983, \mn@doi [\apjl]
  {10.1086/184139}, \href
  {https://ui.adsabs.harvard.edu/abs/1983ApJ...274L...1W} {274, L1}

\makeatother
\end{thebibliography}

%\begin{thebibliography}{99}
%\end{thebibliography}

%%%%%%%%%%%%%%%%%%%%%%%%%%%%%%%%%%%%%%%%%%%%%%%%%%

%%%%%%%%%%%%%%%%% APPENDICES %%%%%%%%%%%%%%%%%%%%%

\appendix
\onecolumn
\counterwithin{figure}{section}
\section{Figures}
\begin{figure}
    \centering
    \includegraphics[width=0.8\linewidth]{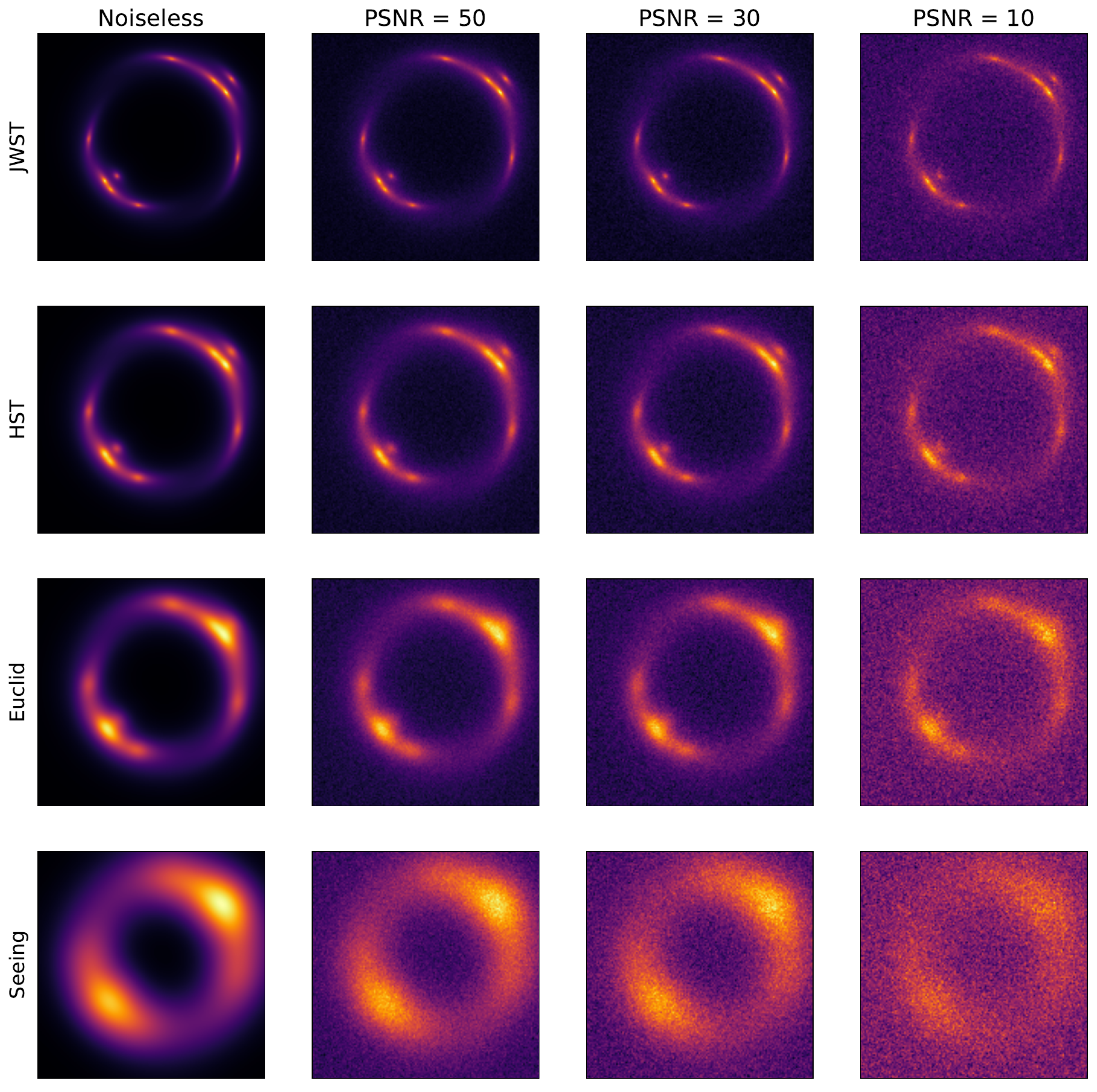}
    \caption{Example of how the noise and FWHM of the PSF change the lenses in our datasets. FWHM's have been chosen so that the resolution are consistent with JWST (0.03"), HST (0.08"), Euclid (0.17"), and natural seeing (0.5")}
    \label{fig:resolution_sample}
\end{figure}

%%%%%%%%%%%%%%%%%%%%%%%%%%%%%%%%%%%%%%%%%%%%%%%%%%

% Don't change these lines
\bsp	% typesetting comment
\label{lastpage}
\end{document}